\documentclass[useAMS,usenatbib]{mn2e}
\usepackage{amsmath}
\usepackage{txfonts}
\usepackage[dvips]{graphicx}
\usepackage{epsfig}
\usepackage{natbib}
\usepackage{subfigure}
\usepackage{multirow}
\usepackage{url}
\bibpunct{(}{)}{;}{a}{}{,}
\usepackage{color}

\title[How robust are predictions of galaxy clustering?]%
{How robust are predictions of galaxy clustering?}

\author[Contreras et al.]
  {
S.~Contreras$^{1,2}$, 
C.~M.~Baugh$^{2}$, 
P.~Norberg$^{2}$, 
N.~Padilla$^{1}$. 
\\
 $^{1}$Departamento de Astronom\'{\i}a y Astrof\'{\i}sica, Pontifica
 Universidad Cat\'olica de Chile, Santiago, Chile.\\
 $^{2}$Institute for Computational Cosmology, Department of Physics, 
Durham University, South Road, Durham, DH1 3LE, UK.
}

\date{Released 2012 Xxxxx XX}
\pagerange{\pageref{firstpage}--\pageref{lastpage}} \pubyear{2012}

\def\LaTeX{L\kern-.36em\raise.3ex\hbox{a}\kern-.15em
    T\kern-.1667em\lower.7ex\hbox{E}\kern-.125emX}

\begin{document}
\label{firstpage}
\maketitle

\begin{abstract}
We use the Millennium Simulation database to compare how different 
versions of the Durham and Munich semi-analytical galaxy formation 
models populate dark matter haloes with galaxies. The models follow 
the same physical processes but differ in how these are implemented. 
All of the models we consider use the Millennium N-body Simulation; 
however, the Durham and Munich groups use independent algorithms 
to construct halo merger histories from the simulation output.   
We compare the predicted halo occupation distributions (HODs) 
and correlation functions for galaxy samples defined by stellar mass, 
cold gas mass and star formation rate. The model predictions 
for the HOD are remarkably similar for samples ranked by stellar mass. 
The predicted bias averaged over pair separations in the range  
$5-25 h^{-1}$Mpc is consistent between models to within 10\%. At small pair 
separations there is a clear difference in the predicted clustering. 
This arises because the Durham models allow some satellite galaxies 
to merge with the central galaxy in a halo when they are still associated 
with resolved dark matter subhaloes. The agreement between the models 
is less good for samples defined by cold gas mass or star formation 
rate, with the spread in predicted galaxy bias reaching 20\% and the 
small scale clustering differing by an order of magnitude, 
reflecting the uncertainty in the modelling of star formation.  
The model predictions in these cases are nevertheless 
qualitatively similar, with a markedly shallower slope for the 
correlation function than is found for stellar mass selected samples 
and with the HOD displaying an asymmetric peak for central galaxies. 
We provide illustrative parametric fits to the HODs predicted by 
the models. Our results reveal the current limitations on how well 
we can predict galaxy bias in a fixed cosmology, which has implications 
for the interpretation of constraints on the physics of galaxy 
formation from galaxy clustering measurements and the ability of 
future galaxy surveys to measure dark energy.
\end{abstract} 


\section{Introduction} 
\label{S_intro}

Galaxy formation is an inefficient process, with only a 
few percent of the available baryons in the Universe 
found in a ``condensed'' state as stars or cold gas 
\citep{Balogh:2001,Cole:2001,McCarthy:2007}. 
The fraction of condensed baryons also varies strongly 
with halo mass, as a result of the interplay between 
the processes which participate in galaxy formation, 
such as gas cooling, star formation, heating of the 
interstellar medium by supernovae and the impact on the 
host galaxy of energy released by AGN 
\citep{Baugh:2006, Benson:2010}. Physical calculations  
of galaxy formation attempt to model all of these processes 
to predict the number of galaxies hosted by a dark matter 
halo and their properties. The aim of this paper is to assess 
how robustly current models can predict the number of galaxies 
which are hosted by dark matter haloes of different mass. 
By focusing on a subset of the predictions possible with 
current models and through selecting galaxies based on their 
intrinsic properties rather using more direct observational 
criteria, we will be able to make a cleaner comparison between 
the models. 

For this comparison we use publicly available galaxy 
catalogues produced by two groups who have independently  
developed semi-analytical models of galaxy formation. 
Semi-analytical models attempt to calculate the fate 
of the baryonic component of the universe, in the context 
of the hierarchical growth of structure in the dark matter. 
These models use differential equations to describe the 
processes listed in the opening paragraph. Often, these 
processes are poorly understood and nonlinear, so the 
equations contain parameters. The values of the parameters 
are set by comparing the model predictions to a selection 
of observational data, and adjusting the parameter values 
until an acceptable match is obtained. Currently, and for 
the foreseeable future, semi-analytical modelling is the 
only technique which can feasibly be used to populate 
large cosmological volumes of dark matter haloes with galaxies 
to obtain predictions for galaxy clustering out to scales 
of several tens of megaparsecs. 

In this comparison we use galaxy formation models which 
have been run using the Millennium N-body simulation 
\citep{Springel:2005}. We consider a range of models run 
by the ``Durham'' and ``Munich'' groups (listed in the next 
section) using the outputs from the dark matter only 
Millennium Simulation. The two groups have their own 
algorithms for constructing merger histories for dark matter 
haloes and different implementations of the physics of 
galaxy formation. Since the models populate the 
same dark matter distribution with galaxies this provides 
an opportunity to look for any systematic differences in the 
predictions for the galaxy content of dark matter haloes. 

The conclusions of this comparison will tell us how 
robust the predictions of current models are, given the uncertainties in the 
underlying physics. This is important for assessing how 
useful measurements of galaxy clustering are for constraining 
the physics of galaxy formation. If the models purport to 
follow the same processes, yet predict different numbers 
of galaxies per halo, then this limits what we can learn 
from clustering at present. As well as improving our understanding 
of physics, modelling galaxy clustering and how it relates 
to the clustering of the underlying dark matter, called 
galaxy bias, is also important for future galaxy surveys 
which aim to measure dark energy \citep{Laureijs:2011,Schlegel:2011}. 
Galaxy bias is a systematic which limits the performance of 
large-scale structure probes of dark energy. If we can model 
bias accurately, then this systematic can be marginalized over.  

This paper is organized as follows. In Section 2 we give a brief 
overview of semi-analytical modelling and state which models we 
are comparing. Section 3 describes some preparatory work for the 
comparison, which involves post-processing of the catalogues and 
describes how we construct our samples. The main results of the 
paper are in Section 4 and our conclusions are presented in 
Section 5. The Appendix describes parametric fits to the halo 
occupation distribution predicted by the models. 

\section{The galaxy formation models} 
\label{S_data}

We compare the predictions of five different semi-analytical 
models of galaxy formation which are publicly available 
from the Millennium Archive in Garching\footnote{\url{http://gavo.mpa-garching.mpg.de/MyMillennium/}} and its mirror 
in Durham\footnote{\url{http://galaxy-catalogue.dur.ac.uk:8080/MyMillennium/}}. The models are all set in the cosmological 
context of the Millennium N-body simulation of the hierarchical 
clustering of matter in a $\Lambda-$CDM cosmology \citep{Springel:2005}. 
The models were produced by two independent groups of researchers, 
and correspond to ``best bet'' models released to the community 
since 2006. Two of the models are generally referred to 
under the label of ``Durham models'' \citep{Bower:2006,Font:2008} and 
will be referred to in plots, respectively, as ``Bower 2006'' 
and ``Font 2008'', and the other three are ``Munich'' models 
\citep{DeLucia:2007,Bertone:2007,Guo:2011}, which will be labelled 
as ``DeLucia2007'', ``Bertone2007'' and ``Guo2011'', respectively.  

The models all aim to follow the main physical processes which 
are believed to be responsible for shaping the formation and 
evolution of galaxies. 
These are: 
(i) the collapse and merging of dark matter haloes; 
(ii) the shock heating and radiative cooling of gas inside dark matter
haloes, leading to the formation of galaxy discs; 
(iii) quiescent star formation in galaxy discs; 
(iv) feedback from supernovae (SNe), accretion of mass onto supermassive 
black holes and from photoionization heating of the intergalactic medium (IGM); 
(v) chemical enrichment of the stars and gas; 
(vi) galaxy mergers driven by dynamical friction within
common dark matter haloes, leading to the formation of
stellar spheroids, which may also trigger bursts of star formation.
However, the implementations of {\it all} of these processes 
differ between the models. These differences even include the 
first step listed above of the construction of halo merger trees 
from the N-body simulation.   
The modelling of the above processes is uncertain and the resulting 
equations often require parameter values to be specified. 
The models differ in how these parameters are set, as the different 
groups assign different importance to reproducing various observational 
datasets. 

It is not our intention to present a comprehensive description of the 
models and their differences. Full details of the models can be found 
in the references given above and in earlier papers by each group. 
The Durham model, {\tt GALFORM}, was introduced by \cite{Cole:2000} and 
extended, for the purposes of the models considered here by 
\cite{Benson:2003}. The Munich model, {\tt LGALAXIES}, was introduced by 
\cite{Springel:2001} and developed in a series of papers \citep{DeLucia:2004, 
Croton:2006, DeLucia:2006}. 

Instead, as we present our results and try to interpret the level 
of agreement between the model predictions, we will discuss various  
components of the models which we believe are responsible 
for any differences. 

\section{Preliminaries: preparation for a comparison between models} 

Having downloaded the galaxy catalogues from the 
Millennium Archive\footnote{The query used is essentially example ``G1'' 
from the ``Mainly galaxies'' demo queries shown on the web page.}, in order to carry out a robust 
and meaningful comparison between the model predictions, 
it is important to account for any differences in definitions 
of properties and to set up well defined samples. 

There are two aspects we need to homogenize for our 
comparison: the definition of mass used to label dark 
matter haloes and the values of galaxy properties 
used to define samples. We deal with each of these in 
turn below. We close this section by discussing a 
relabelling of some of the halo masses that we found  
necessary in the Munich models, due to differences 
in the algorithms used to construct the halo 
merger histories.  

\subsection{Definition of halo mass} 

\begin{figure}
   \includegraphics[width=0.48\textwidth]{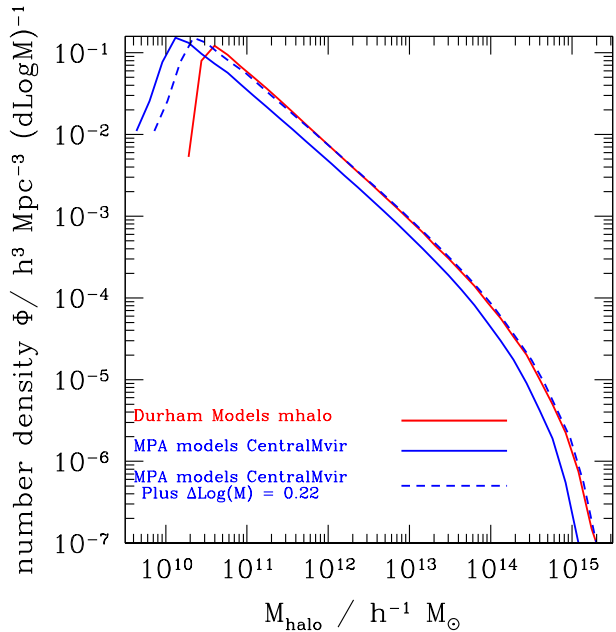}
   \caption{
The mass function of dark matter haloes using the 
original mass ``labels'' obtained from the Millennium 
Archive, shown by the solid blue curve for the Munich models 
(label=centralMvir) and the solid red curve for the 
Durham models (label=mhalo ). 
By rescaling the Munich halo mass labels by 
$\Delta \log_{10} M = 0.22$, as shown by the dashed blue curve, 
the mass functions agree with one another. 
} 
  \label{fig:HMF} 
\end{figure}

The Durham models list halo masses (archive property name ``mhalo'') 
derived from the ``DHalos'' merger tree construction, which 
is described in \cite{Merson:2012} (see also Jiang et~al. in 
preparation).  
The algorithm is designed to ensure that the halo merger trees conserve 
mass and the mass of a branch increases monotonically (or stays the 
same) with time. As a result, the DHalo masses cannot be related in a simple 
way to other measures of mass, such as the number of particles 
identified with a friends-of-friends ({\tt FOF}) percolation group finder. 
The Munich models store ``centralMvir'', which is described as 
the ``virial mass of background ({\tt FOF}) halo containing the galaxy'', 
and is derived from the {\tt FOF} mass. 

The differences in these definitions of halo mass are  
apparent in Fig.~\ref{fig:HMF}, in which we plot the 
mass function of dark matter haloes using the Durham and 
Munich halo mass ``labels''. It is immediately clear 
from this plot that the halo mass labels used by each 
group do not correspond to a simple particle number returned 
by a percolation group finder. The absence of a sharp cut-off in the 
Munich mass function corresponding to the 20 particle limit imposed on 
the list of FOF groups stored is due to how the virial mass 
is estimated from the number of particles that the group finder 
says belong to each halo. 

At the present day the mass functions can be brought into 
remarkably close agreement with one another by rescaling the mass 
in one of the models by a constant factor. We apply this scaling to the Munich 
masses so that afterwards, haloes with 
the same abundance have the same mass\footnote{We note that a similar scaling can be performed to match the halo mass functions at different redshifts. However, the scaling does not work quite so well as it does at $z=0$, and the factor 
required changes with redshift.}. 
We need to make this rescaling as we plot many of our 
comparisons as a function of halo mass. 

\subsection{Galaxy properties}

\begin{figure}
\includegraphics[width=0.48\textwidth]{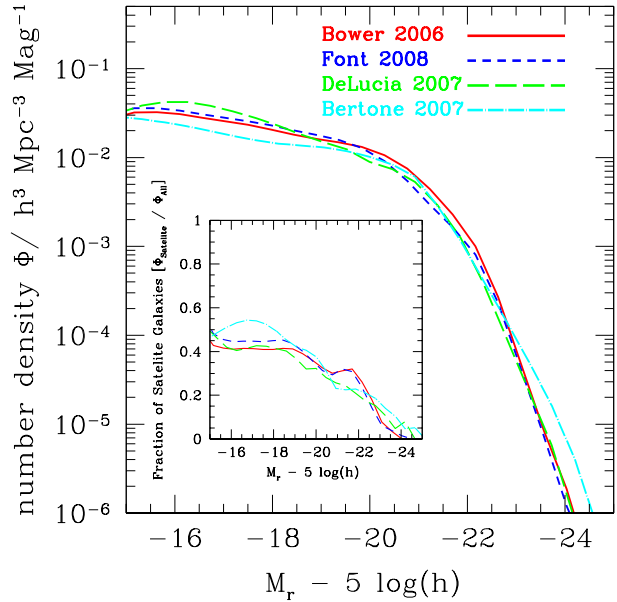}
\includegraphics[width=0.48\textwidth]{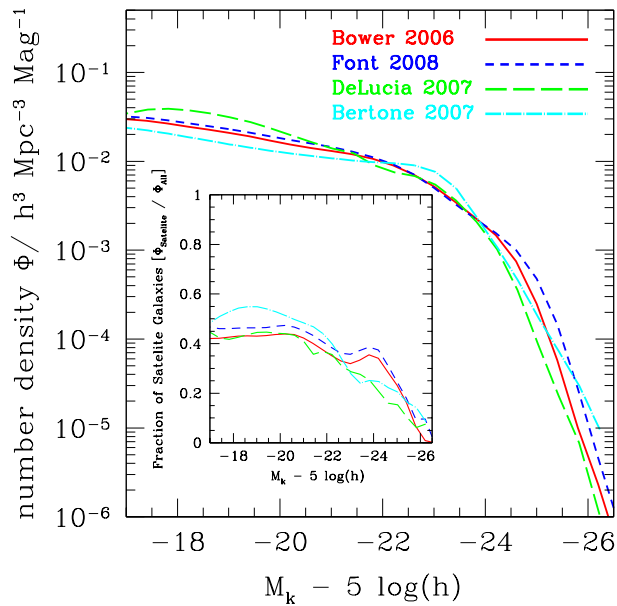}
\caption{
The r-band (top) and K-band (bottom) luminosity functions at $z=0$. 
The model predictions are shown by the different line colours and 
styles, as indicated by the key. Note that the K-band magnitude 
is not available in the Millennium Archive for the Guo et~al. model, 
so we do not show this model in the lower panel. All magnitudes are 
on the Vega scale and include the effects of dust extinction. 
The insets show the fraction of galaxies which are satellites as 
a function of magnitude. The line styles and colours have the same 
meaning as in the main panel.     
}
\label{fig:GLF} 
\end{figure}

%%% SAMs DESCRIPTION %%%-------------------------------------------<T>
\begin{table*}
\begin{center}
\begin{tabular}{ | l | l | l | l | l | l | l | l | l | l |}
  \hline
 &   \multicolumn{3}{|c|}{Density cut 1} & \multicolumn{3}{|c|}{Density cut 2} &\multicolumn{3}{|c|}{ Density cut 3}  \\ 
Abundance & \multicolumn{3}{|c|} {$46.75\times 10^{-3} h^{3} $Mpc$^{-3}$} & \multicolumn{3}{|c|}{$11.77\times 10^{-3} h^{3}$Mpc$^{-3}$}& \multicolumn{3}{|c|}{$0.53\times 10^{-3} h^{3} $Mpc$^{-3}$ } \\ \hline
Model        &$\log(M_{\rm min}^*)$ & $\log(M_{\rm min}^{CG})$ & $ SFR_{\rm min}$&$\log(M_{\rm min}^*)$ & $\log(M_{\rm min}^{CG})$ & $ SFR_{\rm min}$  &$\log(M_{\rm min}^*)$& $\log(M_{\rm min}^{CG})$ & $ SFR_{\rm min}$ \\
  \hline \hline
Bower et al., 2006 & $9.00$ & $9.33$ & $0.070$ & $10.00$ & $9.97$ & $0.65$ & $11.00$ & $10.58$ & $7.91$\\
Font et al., 2008 & $9.02$ & $9.39$ & $0.090 $ & $ 9.99$ & $9.98$ & $ 0.61$  & $11.04$ & $10.58$  & $7.16 $ \\
De Lucia et al., 2007 & $9.17$ & $9.04$ & $0.064 $ & $10.06$ & $9.57$ & $0.58 $  & $11.05$ & $10.33$ & $5.90 $ \\
Guo et al., 2011 & $8.90$ & $9.02$ & $0.0112  $  & $10.06$ & $9.47$ & $0.31 $  & $10.96$ & $10.24$ & $5.14 $ \\
Bertone et al., 2007 & $9.02$ & $8.90$ & $0.006$  & $10.21$ & $9.54$ & $0.66 $  & $11.07$ & $10.40$ & $8.93 $ \\
  \hline
  \end{tabular}
  \caption{
The upper rows give the abundance of galaxies in the three ``density cut'' 
samples used in the paper. The first column below this gives the name 
of the semi-analytical model. Columns 2, 3 and 4 give the cuts applied to 
each model in the logarithm of stellar mass, the logarithm of the cold gas mass 
and the star formation rate, respectively for the highest density sample, 
density cut 1. In all cases the units of mass are $h^{-1}M_{\odot}$ and 
the units of star formation rate are $M_{\odot}yr^{-1}$. 
Columns 5-7 and 8-10 give the analogous cuts for density cuts 2 and 3 
respectively. 
  }
\label{table:description} 
\end{center}
\end{table*}

The luminosity function is the most basic description of the 
galaxy population. As such, reproducing this statistic is a primary 
goal when setting the parameters of a galaxy formation model. 

The present day luminosity functions in the $r-$ and $K-$bands predicted 
by the models are plotted in Fig.~\ref{fig:GLF}. 
Note that we have added $5 \log h$ to the magnitudes stored 
in the Millennium Archive for the Munich models to put them 
onto the same scale as the Durham models. 
The agreement between the model predictions is encouraging. 
To some extent, the model parameters have been selected to reproduce the 
observed luminosity function, and so one might expect the predictions to 
agree even better. The models were not necessarily tuned to explicitly match 
the luminosity functions in these particular bands. The later 
versions of the Munich models put more emphasis on matching 
the inferred stellar mass function. 
Furthermore, other observables are matched at the same time as 
reproducing the luminosity function data, which may have led to compromises 
in the quality of the reproduction of the luminosity function. 
Finally, the parameter values were set by doing comparisons ``by eye'' 
between the model predictions and the data. 
It is not currently possible to provide a definitive list of 
precisely which datasets were used by the various authors 
to set the model parameters, or to specify the priorities 
assigned to the reproduction of different datasets. This 
should become more transparent with future releases of the 
models, following the development of statistical approaches 
to quantify goodness of fit and the weighting of datasets in 
the parameter setting process 
\citep{Henriques:2009, Bower:2010, Henriques:2012}.  

The inset panels in Fig.~\ref{fig:GLF} show the fraction of galaxies 
that are satellites of the central galaxy within each halo 
as a function of magnitude. Again the models show similar 
trends, with just under half of the galaxies being satellites over most 
of the magnitude range plotted, before this value drops steadily  
brightwards of $L_*$. Satellites and centrals are described separately 
in halo occupation distributions, so this will have implications later on. 

\begin{figure*}
   \includegraphics[width=0.43\textwidth, angle=270]{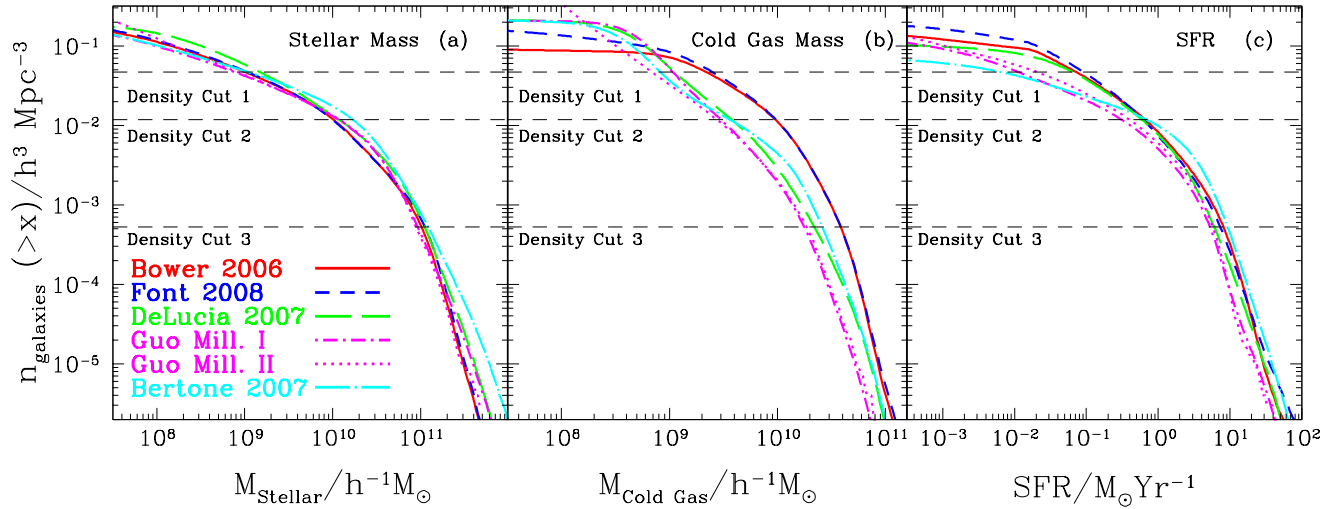}
   \caption{
The cumulative abundance of model galaxies 
ranked by stellar mass (left panel), cold gas mass (middle panel) and 
star formation rate (right panel). As before, the key indicates the 
line colours and style used to represent each model.  
The three horizontal dashed lines in each panel show the three number densities 
(high, intermediate and low) used to define galaxy samples.  
}
   \label{fig:CMF} 
\end{figure*}

\begin{figure*}
\includegraphics[width=1.0\textwidth]{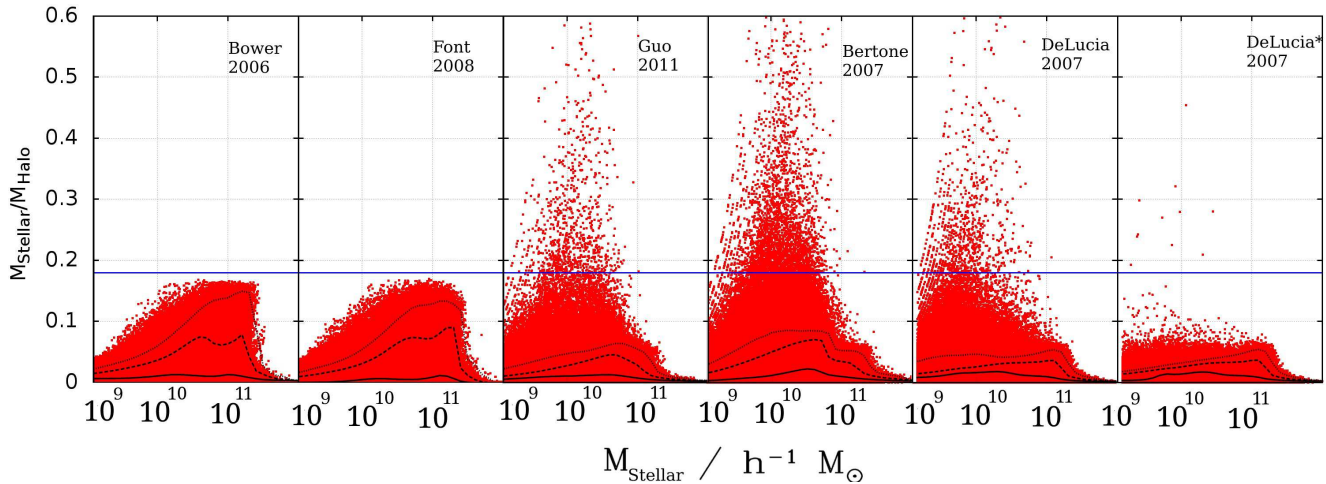}
\caption{
The ratio of stellar mass to the mass of the host dark matter halo 
plotted as a function of stellar mass. The halo masses from the 
Munich models have been globally rescaled as illustrated in Fig.~\ref{fig:HMF}. 
Each panel shows a different model as labelled. 
The horizontal blue solid line shows the maximum possible value of the 
ratio corresponding to the universal baryon fraction being converted 
into stellar mass in one galaxy 
(i.e. with no hot gas, cold gas or satellite galaxies in the halo). 
The black curves show the median (solid), 90th (dashed) and 99th 
(dotted) percentiles for the distribution of predicted ratios. 
The right-most panel shows the DeLucia2007 model after attempting to 
relabel halo masses following a postprocessing of the merger trees 
as explained in the text. 
}
\label{fig:msmh} 
\end{figure*}

\begin{figure*}
   \includegraphics[width=0.96\textwidth]{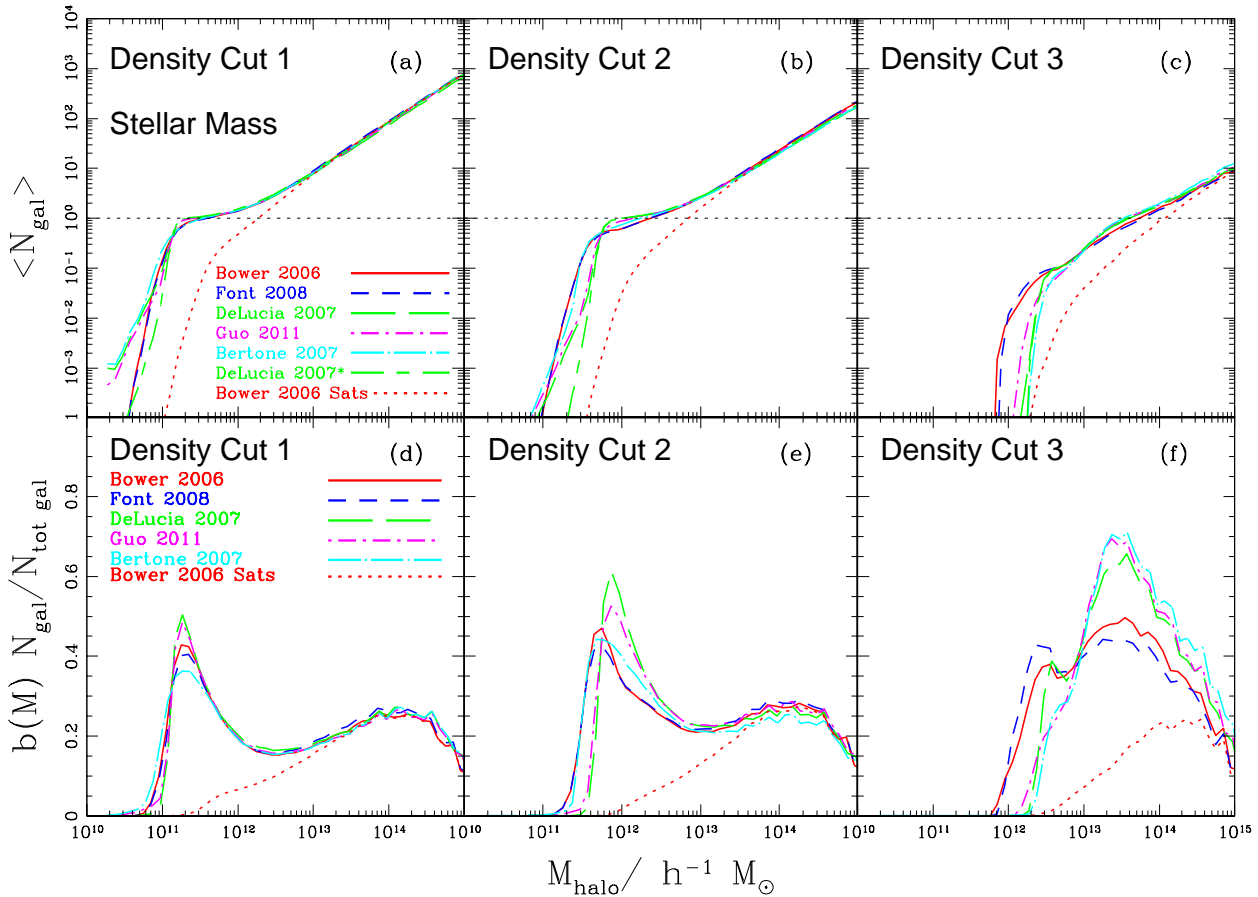}
   \caption{
The top row shows the HODs predicted by the models for the high (left), 
intermediate (middle) and low (right) density samples of galaxies ranked 
in order of descending stellar mass. Different colours correspond to 
different models, as indicated by the key. The DeLucia2007* model corresponds 
to a relabelling of some of the halo masses as described in the text. 
The satellite HOD is shown separately by a red dotted line in the case 
of the Bower2006 model. 
The bottom row shows the contribution to the effective galaxy clustering bias 
as a function of halo mass. Again the dotted red line shows the 
contribution to the bias of satellites in the Bower2006 model. 
Note that the DeLucia2007* model is not plotted in the lower panels 
as it is indistinguishable from the original DeLucia2007 model. 
}
   \label{fig:hodms} 
\end{figure*}

\begin{figure*}
   \centering
   \includegraphics[width=0.43\textwidth, angle=270]{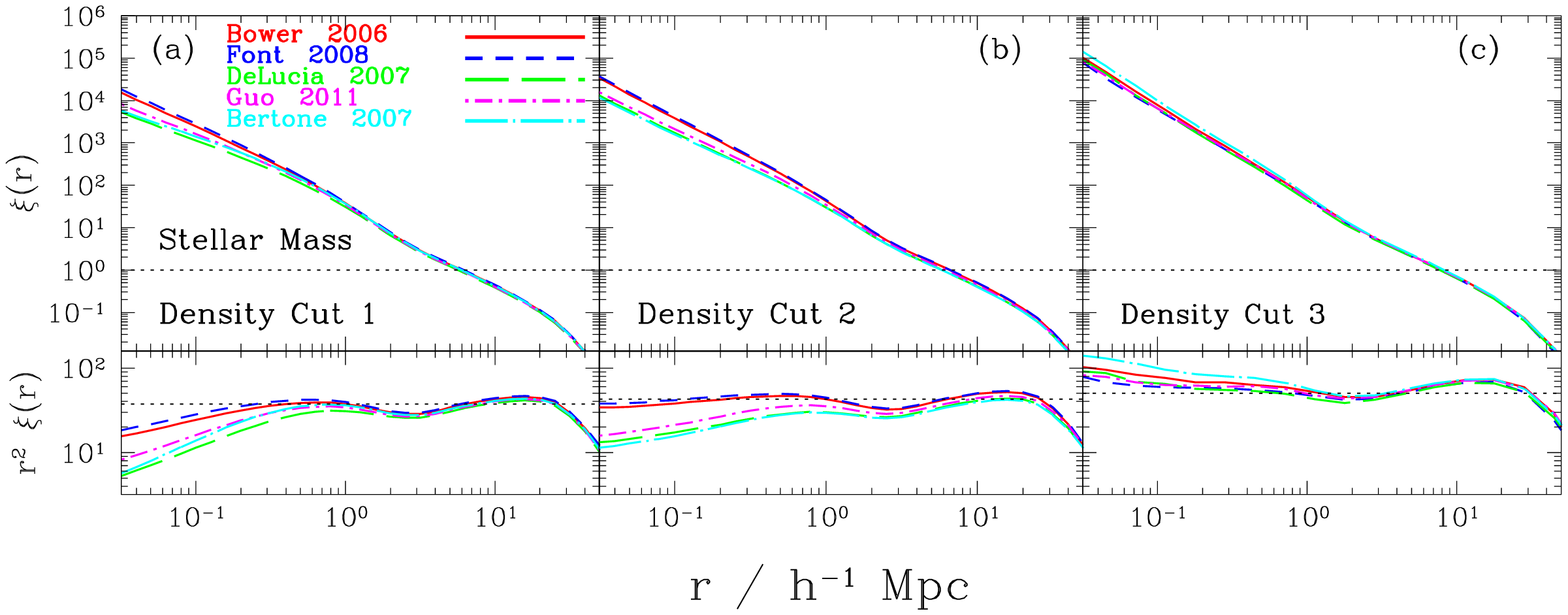}
   \caption{The predicted two-point correlation function in real-space 
for the stellar mass selected samples (high density - left panels; 
intermediate density - middle panels; low density - right panels). 
The upper panels show the correlation function and the lower panels 
show the correlation function multiplied by $r^2$ to emphasize the differences 
between the predictions of the different models. Different colours and 
line styles represent different models, as indicated by the key. 
The dotted lines in the upper panels show $\xi=1$, which can be taken 
as a measure of the correlation length $r_{0}$. The dotted lines 
in the lower panels show $r^{2}_{0}$ for reference, to emphasize the 
departures from a power law.  
   } 
   \label{fig:xims} 
\end{figure*}

When we study the clustering predicted by the models, we will 
use samples defined by intrinsic properties: 
stellar mass, cold gas mass and star formation rate. 
The cumulative abundance of galaxies ranked by each of these 
properties in turn is shown in Fig.~\ref{fig:CMF}. 
There is remarkably close agreement between the distributions 
ranked in terms of stellar mass (left panel of Fig.~\ref{fig:CMF}), 
which is surprising given the differences in the choice 
of stellar initial mass function in the models, which 
means that for a given mass of stars 
made, different amounts of light will be produced, and at least some 
of the models have been specified to reproduce observed luminosity 
functions.  

The model predictions agree less well when galaxies are 
ranked in terms of cold gas. The Durham models predict more galaxies 
for cold gas masses in excess of $10^{9} h^{-1} M_{\odot}$. 
This can be traced to less weight being given to fitting the observed 
gas fractions in spiral galaxies in the Bower2006 and Font2008 models. 
We note that the new treatment of star formation in the Durham models 
introduced by \cite{Lagos:2011b,Lagos:2011a}, which distinguishes 
between atomic and 
molecular hydrogen in the interstellar medium, gives an excellent match 
to the observed HI mass function\footnote{This model is not included in this paper because, at the time of writing, it was not available in the Millennium 
Simulation database. The model will be added early in 2013.}. 
The distributions ranked by SFR are more similar to one another than 
those for cold gas, presumably because in all cases weight was given 
to matching the observed optical colour distributions of galaxies.

\begin{table}
\begin{center}
\begin{tabular}{ | l | c | c | c |}
\hline 
Model &  Stellar  & Cold Gas  & SFR \\ 
      &   mass    &   mass    &     \\
  \hline
Density Cut 2 \\
\hline 
Bower2006  & 40\% & 3\% & 3\% \\
Font2008  & 42\% & 8\% & 10\% \\
DeLucia2007  & 38\% & 6\% & 8\% \\
Bertone2007  & 37\% & 9\% & 14\% \\
Guo2011  & 41\% & 24\% & 19\% \\
\hline 
\hline 
Density Cut 3 \\
\hline 
Bower2006  & 27\% & 2\% & 1\% \\
Font2008  & 26\% & 6\% & 2\% \\
DeLucia2007  & 24\% & 2\% & 4\% \\
Bertone2007  & 27\% & 3\% & 6\% \\
Guo2011  & 26\% & 15\% & 16\% \\
  \hline
  \end{tabular}
  \caption{The percentage 
of galaxies that are satellites in each 
sample. The upper half of the table gives 
the satellite percentages for  
density cut 2 samples and the lower half for density 
cut 3. The first column gives the model label 
and columns 2, 3 and 4 give the satellite percentage 
for samples selected by stellar mass, cold gas mass 
and star formation rate, respectively. 
  }
\label{table:satfrac} 
\end{center}
\end{table}

To ensure we are making a fair comparison between the models, 
we define our galaxy samples by number density rather than by a fixed 
value of a property, such as the stellar mass. Hence, to obtain 
samples which contain the same number of galaxies, slightly different cuts 
on a particular property are applied to each model. The cuts used 
on each property and the number densities of the high, intermediate 
and low density samples are listed in Table 1. This idea of comparing 
galaxy catalogues at a fixed number density was applied by 
\cite{Berlind:2003} and \cite{Zheng:2005} in their HOD analysis of 
galaxies output by a gas dynamic simulation and a semi-analytical model. 
In the Berlind et al. study, the galaxy mass functions output by 
the two galaxy formation models were very different. Nevertheless, by 
comparing samples as a fixed abundance, these authors were able to 
find common features in the HODs. The percentage of galaxies that 
are satellite galaxies is listed in Table~\ref{table:satfrac}. 

The finite resolution of the Millennium simulation means that the 
model predictions are incomplete below some number density. 
To investigate this we plot the results for the Guo et~al. model 
obtained from the Millennium II 
simulation \citep{Boylan-Kolchin:2008}, 
which has a halo mass resolution that is 125 times 
better than that of the Millennium-I run. 
The comparison 
between the galaxy catalogues from the Millennium-I and Millennium-II 
runs shown in Fig.~\ref{fig:CMF} indicates that the Guo et~al. model 
predictions from Millennium-I are robust for stellar masses and star 
formation rates down to our highest number density cut. The cumulative 
distributions agree extremely well from the two simulations for stellar mass. 
In the case of star formation rate, the shapes of the two distributions 
agree quite well, but with a small displacement. For cold gas, the 
distributions are different at low masses. For this reason, we omit 
comparisons corresponding to the first density cut in the case of cold gas 
and star formation rate, which to some extent is controlled by the 
cold gas content. 
Ideally this exercise should be repeated by comparing 
the Millennium-I and Millennium-II predictions for each model in turn, 
as the convergence is likely to be model dependent. However, the Millennium-II 
outputs are not currently available in the database for each model.  

Our methodology of matching the halo mass ``labels'' between models and 
of comparing samples with, by construction, exactly the same number of 
galaxies allows us to focus on differences in the way in which the models 
populate haloes with galaxies. 

\subsection{Further adjustments to the halo masses in the Munich model}

We are almost in a position to compare HODs between models. 
Before we do this, we first carry out a preliminary investigation of 
how galaxy properties correlate with halo mass, by plotting the stellar 
mass to the host halo mass ratio in Fig.~\ref{fig:msmh}. 
This ratio is formed using the mass of the host halo at the present day 
which is the relevant mass for HODs (rather than the subhalo 
mass associated with each galaxy, which is used in subhalo 
abundance matching e.g. \citealt{Simha:2012}). 
The median ratios of stellar mass to halo mass are similar between 
the models, as shown by the solid lines in Fig.~\ref{fig:msmh}. 
The extremes of the distribution and the 
outliers, are, however, different. 

The maximum value of the ratio of stellar mass to host halo mass 
is the universal baryon fraction, since all of the models assume 
that dark matter haloes initially have this baryonic mass associated 
with them. This extreme case corresponds to the presence of a 
single galaxy in the halo with all of the baryons in 
the form of stars, with no hot gas or cold gas. 
The first two panels of Fig.~\ref{fig:msmh} show that the 
Durham models match this expectation with all of the galaxies 
lying below the limit indicated by the horizontal blue dotted 
line. Indeed most galaxies are far away from this line, as 
indicated by the percentile curves, which reflects how inefficient 
galaxy formation is at turning baryons into stars \citep{Cole:2001,Baugh:2006}. 

The Munich models, on the other hand, throw up a small 
number of galaxies (fewer than 1\%) in which the stellar 
mass appears to exceed the universal baryon fraction, in some cases by 
a substantial factor, becoming comparable 
to the host halo mass (in fact in a very small number of cases the 
galaxy stellar masses even exceeds the associated halo mass). 
Closer inspection of the merger histories reveals that these 
galaxies, despite being labelled at the present day as central galaxies, 
reside in haloes that have been tidally stripped in the past. 
The halo spent one or more snapshots orbiting within a larger halo 
during which a substantial amount of mass was lost, leading 
to artificially high stellar mass to halo mass ratios. 
This scenario does not happen by construction with the Durham algorithm 
used to build merger trees (see Merson et~al. 2012 and Jiang et~al., in 
preparation). The stripping of mass from halos is a physical effect. 
However, the extent of the mass stripping will be somewhat dependent 
on the simulation parameters. This ambiguity over the labelling of 
halo masses will have an impact on the shape of the predicted HOD, 
particularly at low halo masses. 

To enable a fair comparison between the predictions between the 
Durham and Munich models, we attempted an approximate correction 
to the Munich halo masses. By examining the galaxy merger trees, 
we identified galaxies which, at the present day, are labelled 
as a central galaxy, but whose host halo was once a satellite 
halo in a more massive structure. We then trace the more massive 
host halo to the present day and assign this halo mass to the 
galaxy instead. The galaxy is relabelled as a satellite and is associated 
with the more massive halo, resulting in the galaxy moving down and 
across in Fig.~\ref{fig:msmh}. 
We have performed this postprocessing only in the case of the 
DeLucia2007 model, which we label as DeLucia2007* in the right-most panel 
of Fig.~\ref{fig:msmh}. This procedure affects $3.6\%$ of the galaxies 
in the DeLucia2007 model, altering the tail of the distribution of 
the stellar mass to halo mass ratio, moving the 99\% line (but not 
the 90\% line) and greatly reduces the number of galaxies whose stellar 
mass exceeds the mass of baryons attached to the associated host halo. 

\section{Results}

The main results of the paper are presented in this section, following 
the preparatory work on the galaxy samples downloaded from the Millennium 
Archive, as outlined in the previous section. In Section 4.1, we compare the 
model predictions for samples defined in terms of stellar mass, looking at the 
HOD and the two-point correlation function. We examine the contribution to the 
correlation function from satellite and central galaxies, and compare the 
radial extent of galaxies in common haloes. A similar study is made 
for samples ranked by cold gas in Section 4.2 and star formation rate in 
Section 4.3. 

\subsection{Stellar mass} 

\begin{figure}
   \centering
   \includegraphics[width=0.48\textwidth, angle=0]{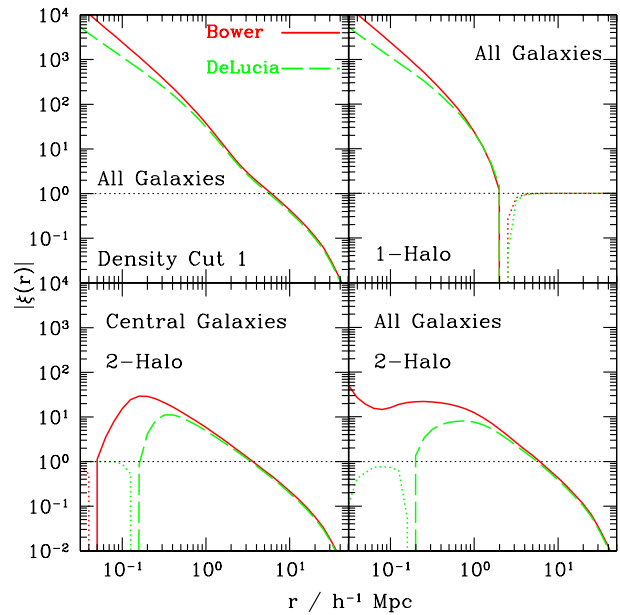}
   \caption{The modulus of the 
two point correlation function predicted by the Bower2006 
and DeLucia2007 models for the highest density stellar mass selected sample. 
The top-left panel shows the full correlation function. The top-right panel 
shows the 1-halo term, corresponding to pairs within the same dark matter 
halo. The lower panels show the 2-halo terms, with the correlation function 
of central-central pairs in the lower-left panel and the full 2-halo term 
in the lower right panel. Dotted curves show where the correlation 
function was negative before taking the modulus. For pair 
separations at which the coloured dotted curves are unity, 
this implies that $\xi = -1$ due to there being no pairs at 
these separations (top-right panel).}
   \label{fig:xi_sample_gL} 
\end{figure}  

 \begin{figure}
   \includegraphics[width=0.48\textwidth]{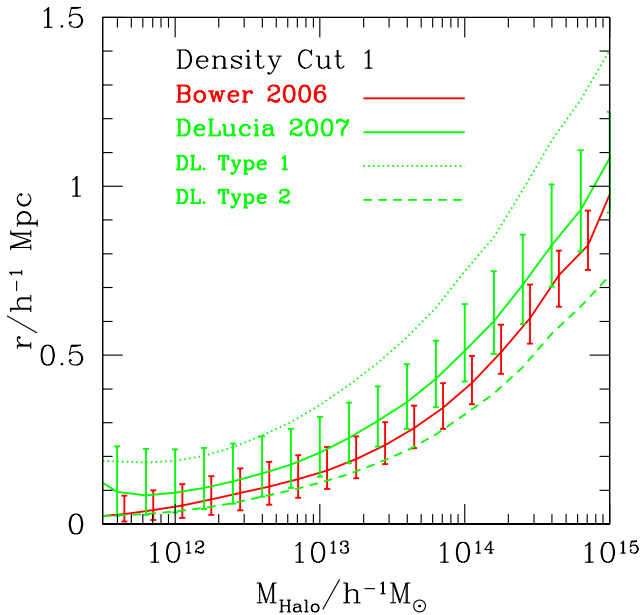}
   \caption{
The median radius of satellites from the central galaxy in a 
dark matter halo, for the highest density sample selected by 
stellar mass, for the Bower2006 and DeLucia2007 models, as 
indicated by the key. The errorbars show the 20-80 percentile 
range of the distribution. The models predict very similar numbers 
of galaxies over this range of halo mass. The galaxy distribution 
has a larger radial extent in the DeLucia model. The median radius for 
type 1 satellites (those which retain an identifiable subhalo; 
dotted line) and type 2 satellites (those for which the associated 
subhalo can no longer be found; dashed line) are also plotted. 
   } 
   \label{fig:radius3} 
\end{figure}

The HODs predicted by the models when galaxies are ranked by 
their stellar mass are plotted in the top panel of 
Fig.~\ref{fig:hodms}. The agreement 
between the model predictions for the high density sample is spectacular 
(left panel of Fig.~\ref{fig:hodms}). 
The Munich models display a slight kink at very low mean numbers of galaxies 
per halo compared to the Durham models. This is the regime in which only 
a tiny fraction of haloes, roughly 1 in 1000, contain a galaxy 
which meets the selection criteria. In the case of DeLucia2007*, 
where we have applied a further correction to a small fraction 
of halo masses as described in the previous section, the agreement with the  
Durham models improves and holds for all masses plotted. 
Here, a galaxy that was originally a central galaxy in a halo 
which has been heavily stripped is now treated as a satellite galaxy 
in a more massive halo. 

The predicted HODs are qualitatively similar for the intermediate and low 
density samples (middle and right panels of Fig.~\ref{fig:hodms}). 
However, in detail the HODs are different, particularly around the halo 
masses at which central galaxies first start to make it into the samples. 
Note that in the lowest density sample (corresponding to the most massive 
galaxies), there is no extended plateau feature corresponding to 
the situation in which all haloes contain a central galaxy that is 
included in the sample. There is a knee in the HOD around a 
halo mass of $\sim 3 \times 10^{12} h^{-1} M_{\odot}$, corresponding to 
the halo mass at which one in ten haloes contains 
a central galaxy which is sufficiently massive to be included 
in the sample. Satellite galaxies make an appreciable contribution 
to the HOD long before the mean number of galaxies per halo reaches unity, 
showing that the canonical HOD model of a step function for centrals which 
reaches unity, plus a power law for satellites, is a poor description of 
the model predictions.  

The variation in the form of the predicted HOD for different 
number densities can be explained in terms of the relative importance 
of AGN feedback which suppresses gas cooling in massive haloes. 
The highest number density sample (ie. corresponding to the lowest 
stellar mass cuts in each model) is dominated by galaxies in haloes 
which are not affected by AGN feedback, since a mean galaxy occupation 
of unity is reached for haloes of mass $\sim 10^{11} h^{-1} M_\odot$. 

For the lower density samples (higher stellar masses), the form 
of the HOD is sensitive to the operation of AGN/radio mode feedback. 
The suppression of cooling in massive haloes changes the slope and 
scatter of the stellar mass - halo mass relation \citep{Gonzalez-Perez:2011}. 
With this feedback mode, the most massive galaxies are not 
necessarily in the most massive haloes. This is particularly true 
in the case of the Durham models, which show more scatter between stellar mass 
and halo mass than is displayed by the Munich models (Contreras et~al. 
in prep.). The consumption of cold gas in a starburst when a disk 
becomes dynamically unstable may also play a role in determining the 
form of the HOD for cold gas selected samples. Unstable disks are more 
common in the Durham models than in the Munich models \citep{Parry:2009}. 

Differences in the HOD curves can seem quite dramatic, particularly 
at low masses where the HOD makes the transition from zero 
galaxies per halo, and at high halo masses, where the HOD is in general 
a power law. However, the HOD does not give the full view of galaxy 
clustering. It is important to bear in mind that the number density of 
dark matter haloes and the clustering bias associated with haloes of 
a given mass also contribute and these quantities change 
significantly across the mass range plotted in Fig.~\ref{fig:hodms}. 
Following \cite{kim:2009}, we show in the bottom panel 
of Fig.~\ref{fig:hodms} the contribution to the effective bias of the galaxy 
sample as a function of halo mass, which takes into account the halo mass 
function and the halo bias. Note that the y-axis in the bottom panels 
is on a linear scale. The high and intermediate density samples 
show two distinct peaks in the contribution to the effective bias, with the low 
mass peak corresponding to central galaxies and the high mass peak to 
satellite galaxies.  For the low density sample (bottom right panel of Fig.~\ref{fig:hodms}), the central galaxy peak is much less distinct and 
satellite galaxies dominate the effective bias. 

The two-point correlation function of the stellar mass selected samples 
is plotted in Fig.~\ref{fig:xims}. At larger pair separations ($1 \le (r/h^{-1}{\rm Mpc}) \le 30 $), the correlation functions predicted by the 
models are remarkably similar, as expected from the similarity in the 
effective bias plotted in Fig.~\ref{fig:hodms}.
The variation in the predicted bias, averaged over pair 
separations of $5 - 25 h^{-1}$Mpc is less than 10\%.  
At small pair separations, for the lowest density sample, the 
correlation functions differ by 10-30\%. 
In the case of the intermediate and high density samples, the extremes of the predictions vary by a factor of 2-3, with more clustering predicted in the Durham models than in the Munich models. 

To gain further insight into the correlation functions 
predicted by the models, we examine the contributions 
from galaxy pairs in the same halo (the 1-halo term in HOD terminology) and 
from different halos (the 2-halo term) in Fig.~\ref{fig:xi_sample_gL}, 
in which we compare the Bower2006 and DeLucia2007 results.  
The top-left panel confirms that the correlation functions 
are remarkably close to one another, except for pair separations 
smaller than $r \sim 1 h^{-1}$Mpc. This difference is dominated 
by the 1-halo term (top-right panel), which itself is largely 
determined by satellite-satellite pairs. In the Bower2006 model 
the 2-halo term makes a small contribution to the amplitude of  
the correlation function at these pair separations, whereas in 
the DeLucia2007, centrals contribute no pairs at these scales. 
This difference can be understood in terms of the DHalos algorithm 
used to build halo merger histories, which breaks up some FOF haloes 
into separate objects, allowing centrals to be found closer 
together than in the Munich models.  

After rescaling the halo masses and considering samples with the 
same number density of galaxies, we find that the Bower2006 
and DeLucia2007 models contain very similar numbers of satellite 
galaxies per halo, as revealed by the close agreement between 
the HODs plotted in Fig.~\ref{fig:hodms} (see Table~\ref{table:satfrac}). 
This implies that overall, the timescale for galaxies to merge 
must be similar in the models. Both models use analytical calculations of 
the time required for a satellite to merge with the central galaxy, 
based on the dynamical friction timescale (see \cite{Lacey:1993}; 
\cite{Cole:2000}). Very similar expressions are used for this 
timescale, with an adjustable parameter chosen to extend the 
timescales in both cases to improve the model predictions for the bright end 
of the galaxy luminosity function. In the Munich models, the 
satellite orbit is followed as long as the associated subhalo 
can be resolved, then the time required for the satellite to 
merge with the central is calculated analytically. In the Durham models, 
the merger timescale is calculated as soon as the galaxy 
becomes a satellite, without any consideration as to whether 
or not the associated subhalo can still be identified.  

The reason for the enhanced small scale clustering in the 
Durham model must therefore be due to a difference in the 
spatial distribution of these satellites within dark matter 
haloes. Fig.~\ref{fig:radius3} confirms this, showing the 
median separation between the central galaxy in a halo and 
its satellites. The median satellite radius is greater in 
the Munich model than in the Durham model. This is readily 
understood from the way in which the merger times are calculated. 
In the Munich model, satellite galaxies whose subhaloes can be 
resolved do not merge with the central galaxy. In the Durham 
model, some fraction of galaxies which are associated with an 
identifiable subhalo will be assumed to have merged, as a result 
of the purely analytical calculation of the merger time. 
The spatial distribution of satellites with resolved subhaloes is 
more extended than the overall distribution of satellites. 
This is clear from Fig.~\ref{fig:radius3} which shows that Type 
1 satellites in the DeLucia2007 model, i.e. those with resolved 
subhaloes, have a larger median radius than satellites whose subhalo 
can no longer be identified (Type 2 satellites). There are around 
three times as many satellites with resolved subhaloes in the 
DeLucia model compared with the Bower model. 

  \begin{figure}
   \centering
   \includegraphics[width=0.5\textwidth]{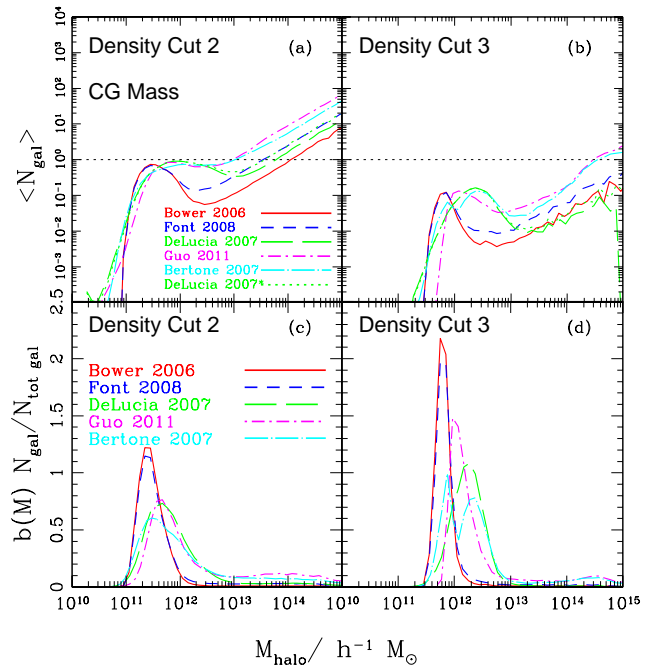}
   \caption{
The upper panels show the predicted HOD for the two lowest density samples 
when galaxies are ranked by their cold gas mass. Different line colours and 
styles refer to the predictions from different models, as indicated by the 
key. The lower panels show the contribution to the effective bias of the 
sample from galaxies in different mass haloes. 
} 
   \label{fig:HODCG} 
\end{figure}    

\begin{figure}
   \centering
   \includegraphics[width=0.31\textwidth, angle=270]{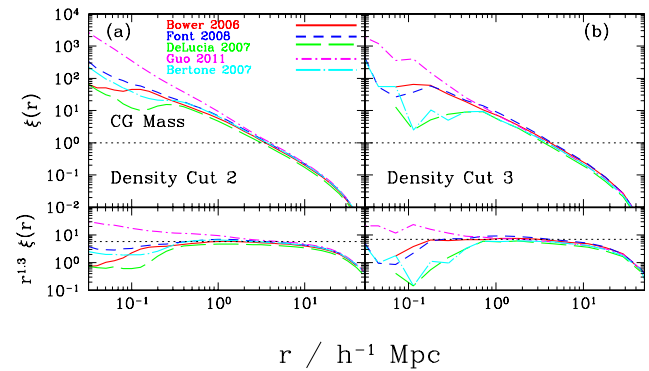}
   \caption{
The predicted two-point correlation function in real-space for 
galaxy samples ranked by their cold gas mass. 
Different line colours and styles refer to the predictions 
from different models, as indicated by the key. The lower panels 
show the correlation functions after multiplying by $r^{1.3}$. 
   } 
   \label{fig:xi_sample_CG} 
\end{figure} 

\subsection{Cold gas mass} 

The models considered in this paper track the total mass in 
cold gas, combining helium and hydrogen, the latter 
in both its atomic and molecular forms. All of this material 
is assumed to be available 
to make stars. With each episode of star formation, some cold 
gas will be ejected from the ISM and perhaps from the dark matter 
halo altogether due to supernovae. More recent models make the 
distinction between molecular and atomic hydrogen using the pressure 
in the mid-plane of the galactic disk and assume that only the molecular 
hydrogen is involved in star formation (see for example \citealt{Lagos:2011a}). 

The comparison presented in Fig.~\ref{fig:CMF} between the 
cumulative cold gas mass function predicted in the Guo et~al. 
model in the Millennium-I and Millennium-II simulations suggests 
that the predictions have not converged for the highest density 
sample, corresponding to the lowest cut in cold gas mass. Hence we 
compare the predicted HODs only for the two lower density cuts in 
Fig.~\ref{fig:HODCG}, which correspond to higher cold gas masses. 
There is a range of HOD predictions. 

The predicted HOD for central galaxies tends to show a peak rather 
than a step. This feature is particularly clear in the lowest 
density sample. This form of the HOD is due to the inclusion of AGN 
feedback in the models, which suppresses gas cooling in massive haloes 
(see the plot of cold gas mass versus halo mass in \citealt{Kim:2011}). 
In the Munich models, the suppression of cooling comes in 
gradually \citep{Croton:2006}, whereas in the Durham models, the 
suppression is total in haloes in which the AGN feedback is operative 
\citep{Bower:2006}. We present a parametric fit to the form of the HOD 
of samples selected by cold gas mass in the Appendix, in which we advocate 
using the nine parameter fit of \cite{Geach:2012} to describe the HOD predicted by the models. 

Satellites make up a smaller fraction of samples selected by cold 
gas mass (see Table~\ref{table:satfrac}). 
The satellite HOD is a power-law as it was in the case of samples 
selected according to their stellar mass, but with a shallower slope 
($\alpha < 1$). There is a substantial range in the amplitude of the 
satellite HODs, which differ by a factor of 5 between the extremes. 
This is consistent with the variation in the percentage of 
galaxies in each sample that are satellites, as listed in 
Table~\ref{table:satfrac}. This is due in part to the way 
in which the model parameters were set. 
The Bower2006 and Font2008 models overpredict 
the observed cold gas mass function \citep{Kim:2011}. The Font2008 
model invokes a revised model for gas cooling in satellites, in which 
the hot gas attached to the satellite is only partially stripped, with  
the fraction of material removed depending upon the orbit of the 
satellite. Hence, gas can still cool onto satellites in the Font et~al. 
model, which explains why the satellite HOD is higher in this model 
than in Bower et~al\footnote{We note that the latest version of the 
Munich model \citep{Guo:2011} includes a similar treatment of gas cooling 
onto satellites to that introduced by Font et~al. However, these authors 
do not present a plot showing how the model predictions compare 
with cold gas data, so we cannot comment on the difference between 
the Guo2011 and Durham models for the cold gas mass function.}. 

The lower panels of Fig.~\ref{fig:HODCG} show that central galaxies 
dominate the contribution to the bias factor for cold gas selected 
samples. The peak in these contributions covers a factor of $\approx 5$ 
in halo mass for the intermediate density sample and an order of magnitude 
for the low density sample. This lack of consensus between the model 
predictions is borne out in the 
predicted correlation functions of cold gas selected samples plotted 
in Fig.~\ref{fig:xi_sample_CG}. On the whole, the correlation function 
for cold gas samples is shallower than that for stellar mass samples, with 
a slope around $-1.3$ rather than $-2$, due to the less important 
role of satellite galaxies in the cold gas samples. There is a 
spread of a factor of $\approx 1.3$ in the clustering 
amplitude around $r \sim 10 h^{-1}$Mpc and of around an 
order of magnitude or more at the smallest pair 
separations plotted. 

\begin{figure}
   \centering
   \includegraphics[width=0.5\textwidth]{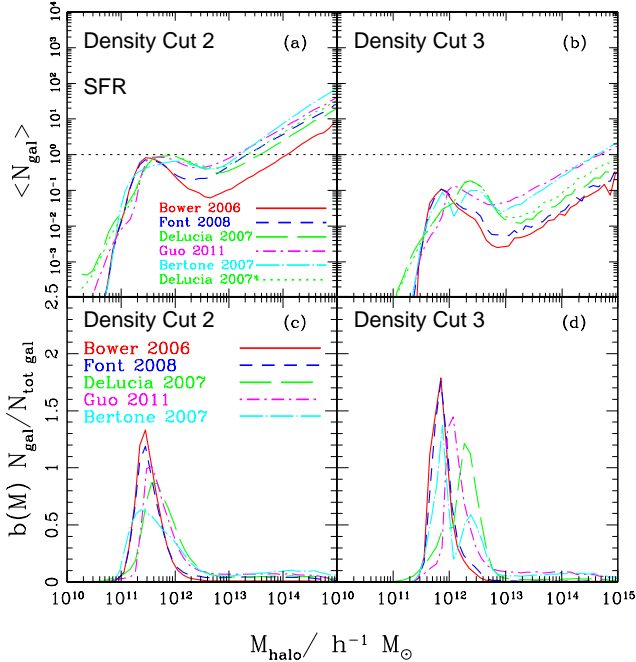}
   \caption{
The upper panels show the predicted HOD for the two lowest density samples 
when galaxies are ranked by their star formation rate. 
Different line colours and styles refer to the predictions from 
different models, as indicated by the key. The lower panels show 
the contribution to the effective bias of the 
sample from galaxies in different mass haloes. 
   } 
   \label{fig:HODSFR} 
\end{figure} 

\begin{figure}
   \centering
   \includegraphics[width=0.31\textwidth, angle=270]{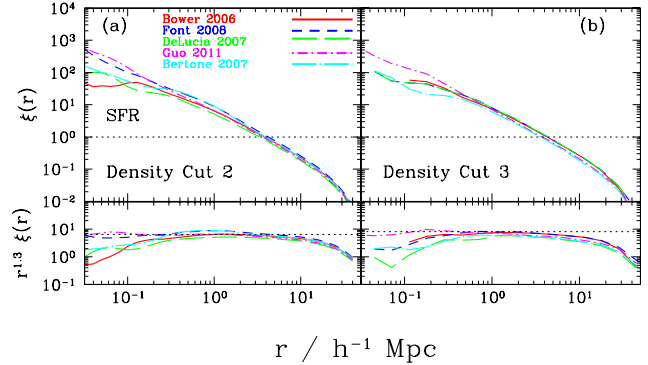}
   \caption{
The predicted two-point correlation function in real-space for 
samples ranked by their star formation rate. 
Different line colours and styles refer to the predictions 
from different models, as indicated by the key. The lower panels 
show the correlation functions after multiplying by $r^{1.3}$. 
}
   \label{fig:xi_sample_SFR} 
\end{figure}

\subsection{Star formation rate} 

Finally, we repeat the analysis of the previous section using 
samples ranked by the global star formation rate in the galaxy, 
which is relevant to observational samples selected by emission 
line strength or rest-frame UV luminosity. The predicted HODs 
plotted in Fig.~\ref{fig:HODSFR} are very similar to those for the 
cold gas samples plotted in Fig.~\ref{fig:HODCG}. In the Bower2006 
and Font2008 models there is a direct connection between the 
total cold gas mass and the star formation rate. In the Munich 
models, only the gas above a surface density threshold is assumed to 
participate in star formation. The correlation functions plotted 
in Fig.~\ref{fig:xi_sample_SFR} also show a shallower slope than 
those obtained for the stellar mass selected samples. The predicted 
clustering shows a similar spread to that displayed for cold gas 
selected samples.

\section{Conclusions} 
\label{s_conclusus} 

We have tested the robustness of the predictions of 
semi-analytical models of galaxy formation for the 
clustering of galaxies. This project was made possible 
through the Millennium Simulation database, which allows 
the outputs of the models to be downloaded and analyzed 
by the astronomical community. 

The models tested were produced by the ``Durham'' 
and ``Munich'' groups, and represent different versions 
of their ``best bet'' models when originally published. 
These codes attempt to model the fate of the baryonic 
component of the universe. The underlying physics is 
complex and still poorly understood despite much 
progress over the past twenty years. Although the two 
groups developed their approaches starting from 
\cite{White:1991}, the implementations of the different processes 
which influence galaxy formation are quite different, as is the  
emphasis on which observations should be reproduced by 
the models in order to set the values of the model parameters. 
The calculations are set in the Millennium N-body 
simulation of \cite{Springel:2005}. However, all aspects 
of galaxy formation modelling beyond the distribution of the 
dark matter, including the construction of merger histories 
for halos, are independent. 

Given the uncertainty in the modelling of galaxy formation, 
the availability of model outputs from the same N-body 
simulation gives us an opportunity to perform a direct 
comparison of how the models populate halos with galaxies. 
We have carried out this comparison by looking at the model 
predictions for the halo occupation distribution (which it 
should be stressed is a model {\it output}) and the two-point 
correlation function. We compare samples at fixed abundances, 
high, medium and low, to account for any differences in the 
distributions of galaxy properties which persist between 
the model predictions. We also use a common definition of 
halo mass, again by referring to the abundance of haloes. 
We look at samples defined by intrinsic properties: stellar 
mass, cold gas mass and star formation rate. 

In the case of samples selected by their stellar mass, 
the agreement in the predicted HODs is remarkable, 
particularly as galaxy clustering is in general not one of the 
observables used to specify the model parameters. 
The two-halo contributions to the correlation function 
are the same in different models, which means that, 
under the conditions of the controlled comparison carried 
out here, the prediction for the asymptotic galaxy bias 
is robust. 

The one-halo term, i.e. the contribution from pairs in 
the same dark matter halo, is different. Again, under the conditions of 
our comparison, there is little difference in the number 
of satellite galaxies in each halo which implies that the 
calculation of the timescale required for a galaxy merger 
to take place is similar in the two sets of models. The difference 
lies in which galaxies are considered for mergers. In the Durham 
models, any satellite can merge provided that the dynamical 
friction timescale is sufficiently short for the merger to have 
time to take place. In the Munich models, only those satellites 
which are hosted by subhaloes which can no longer be identified 
can merge. The radial distribution of subhaloes is more extended 
than the distribution of dark matter \citep{Angulo:2009}, 
so this naturally leads to larger 1-halo pair separations 
in the Munich models. It is fair to argue that here the Munich 
approach is more reasonable, though this depends on how well 
the subhalo finder works. 

\cite{Budzynski:2012} compared the radial density profile of 
galaxies in clusters using a subset of the semi-analytical models 
considered here, and argued that, for their selection, the Durham 
models produced longer merger timescales for satellites than in 
the Munich models, to explain a steeper inner profile. Our conclusion 
is different, as we argue that the merger timescales are similar but that 
the effective radial distribution of galaxies is different. This 
difference in interpretation could be due, however, to the way 
in which galaxy samples are compiled in both studies. 

The agreement between the models is less good when looking 
at samples selected by cold gas mass and star formation rate. 
In these cases the models give qualitatively similar 
predictions which differ in detail. 
There is a small spread in the asymptotic bias (a factor of $\sim 1.2$) 
and a large difference (order of magnitude) in the small scale 
correlation function. One notable difference from the samples 
selected by stellar mass is the form of the HOD predicted for 
cold gas and star formation rate samples. In the latter two 
cases the HOD for central galaxies has an asymmetrical peak, 
rather than the worn step function seen for stellar mass samples. 

The conclusion of our study is that for some galaxy properties, 
different models of galaxy formation give encouragingly similar 
predictions for galaxy clustering. This means that there is some 
hope of understanding galaxy bias in terms of the implications 
for galaxy formation physics, and from the point of view of removing 
it as a systematic in dark energy probes. However, this depends 
on the sample selection, and for surveys which target emission lines, 
the current predictions agree less well. Also, the absolute value 
of the bias predicted by the models may differ if an observational 
selection is applied to the catalogues, as would be done in a mock 
catalogue, rather than performing the controlled test we carried out 
of comparing samples at a fixed number density. 
It is also important to bear in mind that we have tested how closely 
the models predict galaxy bias in a single cosmology. In reality, we 
do not know the true underlying cosmology in the real Universe, and this 
will further complicate the attempt to extract the underlying clustering 
of the mass. As the modelling of the gas content and star formation rate 
in galaxies improves \citep{Lagos:2011a}, hopefully the predictions 
for samples selected by e.g. emission line properties will become more 
robust.

\section*{Acknowledgements}
This work would have been much more difficult without 
the efforts of Gerard Lemson and colleagues at the 
German Astronomical Virtual Observatory in setting up 
the Millennium Simulation database in Garching, 
and John Helly and Lydia Heck in setting up the mirror at Durham. 
We acknowledge helpful conversations with the participants 
in the GALFORM lunches in Durham. This work was partially 
supported by ``Centro de Astronom\'\i a y Tecnolog\'\i 
as Afines" BASAL PFB-06. NP was supported by Fondecyt 
Regular \#1110328. We acknowledge support from the European 
Commission's Framework Programme 7, through the Marie Curie 
International Research 
Staff Exchange Scheme LACEGAL (PIRSES-GA-2010-269264).
PN acknowledges the support of the Royal Society through the award 
of a University Research Fellowship and the European Research 
Council, through receipt of a Starting Grant (DEGAS-259586). 
The calculations for this paper were performed on the 
ICC Cosmology Machine, which is part of the DiRAC-2 
Facility jointly funded by STFC, the Large Facilities 
Capital Fund of BIS, and Durham University and on the Geryon
cluster at the Centro de Astro-Ingenieria at Universidad 
Cat\'{o}lica de Chile.

\bibliographystyle{mn2e}
\bibliography{Biblio}

\begin{thebibliography}{40}
\expandafter\ifx\csname natexlab\endcsname\relax\def\natexlab#1{#1}\fi

\bibitem[{{Angulo} {et~al}\mbox{.}(2009){Angulo}, {Lacey}, {Baugh}, \&
  {Frenk}}]{Angulo:2009}
{Angulo} R.~E., {Lacey} C.~G., {Baugh} C.~M., {Frenk} C.~S., 2009, \mnras, 399,
  983

\bibitem[{{Balogh} {et~al}\mbox{.}(2001){Balogh}, {Pearce}, {Bower}, \&
  {Kay}}]{Balogh:2001}
{Balogh} M.~L., {Pearce} F.~R., {Bower} R.~G., {Kay} S.~T., 2001, \mnras, 326,
  1228

\bibitem[{{Baugh}(2006)}]{Baugh:2006}
{Baugh} C.~M., 2006, Reports on Progress in Physics, 69, 3101

\bibitem[{{Benson}(2010)}]{Benson:2010}
{Benson} A.~J., 2010, \physrep, 495, 33

\bibitem[{{Benson} {et~al}\mbox{.}(2003){Benson}, {Bower}, {Frenk}, {Lacey},
  {Baugh}, \& {Cole}}]{Benson:2003}
{Benson} A.~J., {Bower} R.~G., {Frenk} C.~S., {Lacey} C.~G., {Baugh} C.~M.,
  {Cole} S., 2003, \apj, 599, 38

\bibitem[{{Berlind} {et~al}\mbox{.}(2003){Berlind}, {Weinberg}, {Benson},
  {Baugh}, {Cole}, {Dav{\'e}}, {Frenk}, {Jenkins}, {Katz}, \&
  {Lacey}}]{Berlind:2003}
{Berlind} A.~A. {et~al.}, 2003, \apj, 593, 1

\bibitem[{{Bertone} {et~al}\mbox{.}(2007){Bertone}, {De Lucia}, \&
  {Thomas}}]{Bertone:2007}
{Bertone} S., {De Lucia} G., {Thomas} P.~A., 2007, \mnras, 379, 1143

\bibitem[{{Bower} {et~al}\mbox{.}(2006){Bower}, {Benson}, {Malbon}, {Helly},
  {Frenk}, {Baugh}, {Cole}, \& {Lacey}}]{Bower:2006}
{Bower} R.~G., {Benson} A.~J., {Malbon} R., {Helly} J.~C., {Frenk} C.~S.,
  {Baugh} C.~M., {Cole} S., {Lacey} C.~G., 2006, \mnras, 370, 645

\bibitem[{{Bower} {et~al}\mbox{.}(2010){Bower}, {Vernon}, {Goldstein},
  {Benson}, {Lacey}, {Baugh}, {Cole}, \& {Frenk}}]{Bower:2010}
{Bower} R.~G., {Vernon} I., {Goldstein} M., {Benson} A.~J., {Lacey} C.~G.,
  {Baugh} C.~M., {Cole} S., {Frenk} C.~S., 2010, \mnras, 407, 2017

\bibitem[{{Boylan-Kolchin} {et~al}\mbox{.}(2008){Boylan-Kolchin}, {Ma}, \&
  {Quataert}}]{Boylan-Kolchin:2008}
{Boylan-Kolchin} M., {Ma} C.-P., {Quataert} E., 2008, \mnras, 383, 93

\bibitem[{{Budzynski} {et~al}\mbox{.}(2012){Budzynski}, {Koposov}, {McCarthy},
  {McGee}, \& {Belokurov}}]{Budzynski:2012}
{Budzynski} J.~M., {Koposov} S.~E., {McCarthy} I.~G., {McGee} S.~L.,
  {Belokurov} V., 2012, \mnras, 423, 104

\bibitem[{{Cole} {et~al}\mbox{.}(2000){Cole}, {Lacey}, {Baugh}, \&
  {Frenk}}]{Cole:2000}
{Cole} S., {Lacey} C.~G., {Baugh} C.~M., {Frenk} C.~S., 2000, \mnras, 319, 168

\bibitem[{{Cole} {et~al}\mbox{.}(2001){Cole}, {Norberg}, {Baugh}, {Frenk},
  {Bland-Hawthorn}, {Bridges}, {Cannon}, {Colless}, {Collins}, {Couch},
  {Cross}, {Dalton}, {De Propris}, {Driver}, {Efstathiou}, {Ellis},
  {Glazebrook}, {Jackson}, {Lahav}, {Lewis}, {Lumsden}, {Maddox}, {Madgwick},
  {Peacock}, {Peterson}, {Sutherland}, \& {Taylor}}]{Cole:2001}
{Cole} S. {et~al.}, 2001, \mnras, 326, 255

\bibitem[{{Cooray} \& {Sheth}(2002)}]{Cooray:2002}
{Cooray} A., {Sheth} R., 2002, \physrep, 372, 1

\bibitem[{{Croton} {et~al}\mbox{.}(2006){Croton}, {Springel}, {White}, {De
  Lucia}, {Frenk}, {Gao}, {Jenkins}, {Kauffmann}, {Navarro}, \&
  {Yoshida}}]{Croton:2006}
{Croton} D.~J. {et~al.}, 2006, \mnras, 365, 11

\bibitem[{{De Lucia} \& {Blaizot}(2007)}]{DeLucia:2007}
{De Lucia} G., {Blaizot} J., 2007, \mnras, 375, 2

\bibitem[{{De Lucia} {et~al}\mbox{.}(2004){De Lucia}, {Kauffmann}, \&
  {White}}]{DeLucia:2004}
{De Lucia} G., {Kauffmann} G., {White} S.~D.~M., 2004, \mnras, 349, 1101

\bibitem[{{De Lucia} {et~al}\mbox{.}(2006){De Lucia}, {Springel}, {White},
  {Croton}, \& {Kauffmann}}]{DeLucia:2006}
{De Lucia} G., {Springel} V., {White} S.~D.~M., {Croton} D., {Kauffmann} G.,
  2006, \mnras, 366, 499

\bibitem[{{Font} {et~al}\mbox{.}(2008){Font}, {Bower}, {McCarthy}, {Benson},
  {Frenk}, {Helly}, {Lacey}, {Baugh}, \& {Cole}}]{Font:2008}
{Font} A.~S. {et~al.}, 2008, \mnras, 289, 1619

\bibitem[{{Geach} {et~al}\mbox{.}(2012){Geach}, {Sobral}, {Hickox}, {Wake},
  {Smail}, {Best}, {Baugh}, \& {Stott}}]{Geach:2012}
{Geach} J.~E., {Sobral} D., {Hickox} R.~C., {Wake} D.~A., {Smail} I., {Best}
  P.~N., {Baugh} C.~M., {Stott} J.~P., 2012, \mnras, 426, 679

\bibitem[{{Gonzalez-Perez} {et~al}\mbox{.}(2011){Gonzalez-Perez}, {Baugh},
  {Lacey}, \& {Kim}}]{Gonzalez-Perez:2011}
{Gonzalez-Perez} V., {Baugh} C.~M., {Lacey} C.~G., {Kim} J.-W., 2011, \mnras,
  417, 517

\bibitem[{{Guo} {et~al}\mbox{.}(2011){Guo}, {White}, {Boylan-Kolchin}, {De
  Lucia}, {Kauffmann}, {Lemson}, {Li}, {Springel}, \& {Weinmann}}]{Guo:2011}
{Guo} Q. {et~al.}, 2011, \mnras, 413, 101

\bibitem[{{Henriques} {et~al}\mbox{.}(2012){Henriques}, {White}, {Thomas},
  {Angulo}, {Guo}, {Lemson}, \& {Springel}}]{Henriques:2012}
{Henriques} B., {White} S., {Thomas} P., {Angulo} R., {Guo} Q., {Lemson} G.,
  {Springel} V., 2012, ArXiv e-prints, arXiv1212.1717

\bibitem[{{Henriques} {et~al}\mbox{.}(2009){Henriques}, {Thomas}, {Oliver}, \&
  {Roseboom}}]{Henriques:2009}
{Henriques} B.~M.~B., {Thomas} P.~A., {Oliver} S., {Roseboom} I., 2009, \mnras,
  396, 535

\bibitem[{{Kim} {et~al}\mbox{.}(2011){Kim}, {Baugh}, {Benson}, {Cole}, {Frenk},
  {Lacey}, {Power}, \& {Schneider}}]{Kim:2011}
{Kim} H.-S., {Baugh} C.~M., {Benson} A.~J., {Cole} S., {Frenk} C.~S., {Lacey}
  C.~G., {Power} C., {Schneider} M., 2011, \mnras, 414, 2367

\bibitem[{{Kim} {et~al}\mbox{.}(2009){Kim}, {Baugh}, {Cole}, {Frenk}, \&
  {Benson}}]{kim:2009}
{Kim} H.-S., {Baugh} C.~M., {Cole} S., {Frenk} C.~S., {Benson} A.~J., 2009,
  \mnras, 400, 1527

\bibitem[{{Lacey} \& {Cole}(1993)}]{Lacey:1993}
{Lacey} C., {Cole} S., 1993, \mnras, 262, 627

\bibitem[{{Lagos} {et~al}\mbox{.}(2011{\natexlab{a}}){Lagos}, {Baugh}, {Lacey},
  {Benson}, {Kim}, \& {Power}}]{Lagos:2011b}
{Lagos} C. D.~P., {Baugh} C.~M., {Lacey} C.~G., {Benson} A.~J., {Kim} H.-S.,
  {Power} C., 2011{\natexlab{a}}, \mnras, 418, 1649

\bibitem[{{Lagos} {et~al}\mbox{.}(2011{\natexlab{b}}){Lagos}, {Lacey}, {Baugh},
  {Bower}, \& {Benson}}]{Lagos:2011a}
{Lagos} C. D.~P., {Lacey} C.~G., {Baugh} C.~M., {Bower} R.~G., {Benson} A.~J.,
  2011{\natexlab{b}}, \mnras, 416, 1566

\bibitem[{{Laureijs} {et~al}\mbox{.}(2011){Laureijs}, {Amiaux}, {Arduini},
  {Augu{\`e}res}, {Brinchmann}, {Cole}, {Cropper}, {Dabin}, {Duvet}, {Ealet},
  \& et~al.}]{Laureijs:2011}
{Laureijs} R. {et~al.}, 2011, ArXiv e-prints, arXiv1110.3193

\bibitem[{{McCarthy} {et~al}\mbox{.}(2007){McCarthy}, {Bower}, \&
  {Balogh}}]{McCarthy:2007}
{McCarthy} I.~G., {Bower} R.~G., {Balogh} M.~L., 2007, \mnras, 377, 1457

\bibitem[{{Merson} {et~al}\mbox{.}(2012){Merson}, {Baugh}, {Helly},
  {Gonzalez-Perez}, {Cole}, {Bielby}, {Norberg}, {Frenk}, {Benson}, {Bower},
  {Lacey}, \& {Lagos}}]{Merson:2012}
{Merson} A.~I. {et~al.}, 2012, \mnras, 342

\bibitem[{{Parry} {et~al}\mbox{.}(2009){Parry}, {Eke}, \& {Frenk}}]{Parry:2009}
{Parry} O.~H., {Eke} V.~R., {Frenk} C.~S., 2009, \mnras, 396, 1972

\bibitem[{{Schlegel} {et~al}\mbox{.}(2011){Schlegel}, {Abdalla}, {Abraham},
  {Ahn}, {Allende Prieto}, {Annis}, {Aubourg}, {Azzaro}, {Baltay}, {Baugh},
  {Bebek}, {Becerril}, {Blanton}, {Bolton}, {Bromley}, {Cahn}, {Carton},
  {Cervantes-Cota}, {Chu}, {Cortes}, {Dawson}, {Dey}, {Dickinson}, {Diehl},
  {Doel}, {Ealet}, {Edelstein}, {Eppelle}, {Escoffier}, {Evrard}, {Faccioli},
  {Frenk}, {Geha}, {Gerdes}, {Gondolo}, {Gonzalez-Arroyo}, {Grossan},
  {Heckman}, {Heetderks}, {Ho}, {Honscheid}, {Huterer}, {Ilbert}, {Ivans},
  {Jelinsky}, {Jing}, {Joyce}, {Kennedy}, {Kent}, {Kieda}, {Kim}, {Kim},
  {Kneib}, {Kong}, {Kosowsky}, {Krishnan}, {Lahav}, {Lampton}, {LeBohec}, {Le
  Brun}, {Levi}, {Li}, {Liang}, {Lim}, {Lin}, {Linder}, {Lorenzon}, {de la
  Macorra}, {Magneville}, {Malina}, {Marinoni}, {Martinez}, {Majewski},
  {Matheson}, {McCloskey}, {McDonald}, {McKay}, {McMahon}, {Menard},
  {Miralda-Escude}, {Modjaz}, {Montero-Dorta}, {Morales}, {Mostek}, {Newman},
  {Nichol}, {Nugent}, {Olsen}, {Padmanabhan}, {Palanque-Delabrouille}, {Park},
  {Peacock}, {Percival}, {Perlmutter}, {Peroux}, {Petitjean}, {Prada},
  {Prieto}, {Prochaska}, {Reil}, {Rockosi}, {Roe}, {Rollinde}, {Roodman},
  {Ross}, {Rudnick}, {Ruhlmann-Kleider}, {Sanchez}, {Sawyer}, {Schimd},
  {Schubnell}, {Scoccimaro}, {Seljak}, {Seo}, {Sheldon}, {Sholl},
  {Shulte-Ladbeck}, {Slosar}, {Smith}, {Smoot}, {Springer}, {Stril}, {Szalay},
  {Tao}, {Tarle}, {Taylor}, {Tilquin}, {Tinker}, {Valdes}, {Wang}, {Wang},
  {Weaver}, {Weinberg}, {White}, {Wood-Vasey}, {Yang}, {Yeche}, {Zakamska},
  {Zentner}, {Zhai}, \& {Zhang}}]{Schlegel:2011}
{Schlegel} D. {et~al.}, 2011, ArXiv e-prints, arXiv1106.1706

\bibitem[{{Simha} {et~al}\mbox{.}(2012){Simha}, {Weinberg}, {Dav{\'e}},
  {Fardal}, {Katz}, \& {Oppenheimer}}]{Simha:2012}
{Simha} V., {Weinberg} D.~H., {Dav{\'e}} R., {Fardal} M., {Katz} N.,
  {Oppenheimer} B.~D., 2012, \mnras, 423, 3458

\bibitem[{{Springel} {et~al}\mbox{.}(2005){Springel}, {White}, {Jenkins},
  {Frenk}, {Yoshida}, {Gao}, {Navarro}, {Thacker}, {Croton}, {Helly},
  {Peacock}, \& {Cole}}]{Springel:2005}
{Springel} V. {et~al.}, 2005, \nat, 435, 629

\bibitem[{{Springel} {et~al}\mbox{.}(2001){Springel}, {White}, {Tormen}, \&
  {Kauffmann}}]{Springel:2001}
{Springel} V., {White} S. D.~M., {Tormen} G., {Kauffmann} G., 2001, \mnras,
  328, 726

\bibitem[{{White} \& {Frenk}(1991)}]{White:1991}
{White} S.~D.~M., {Frenk} C.~S., 1991, \apj, 379, 52

\bibitem[{{Zehavi} {et~al}\mbox{.}(2005){Zehavi}, {Zheng}, {Weinberg},
  {Frieman}, {Berlind}, {Blanton}, {Scoccimarro}, {Sheth}, {Strauss}, {Kayo},
  {Suto}, {Fukugita}, {Nakamura}, {Bahcall}, {Brinkmann}, {Gunn}, {Hennessy},
  {Ivezi{\'c}}, {Knapp}, {Loveday}, {Meiksin}, {Schlegel}, {Schneider},
  {Szapudi}, {Tegmark}, {Vogeley}, {York}, \& {SDSS
  Collaboration}}]{Zehavi:2005}
{Zehavi} I. {et~al.}, 2005, \apj, 630, 1

\bibitem[{{Zheng} {et~al}\mbox{.}(2005){Zheng}, {Berlind}, {Weinberg},
  {Benson}, {Baugh}, {Cole}, {Dav{\'e}}, {Frenk}, {Katz}, \&
  {Lacey}}]{Zheng:2005}
{Zheng} Z. {et~al.}, 2005, \apj, 633, 791

\end{thebibliography}

\appendix

\section{Parametric fits to the HODs predicted by the 
semi-analytical models}

\begin{table}
\begin{center}
\begin{tabular}{ | l | l | l | l | l | l |}
\hline
Model & $\alpha$ & $ M_{\rm cut}$       & $M_{\rm min}$         & $M_{1}$ & $\sigma_{\log M}$ \\ 
      &          & $ (h^{-1}M_{\odot})$ & $(h^{-1}M_{\odot})$   &  $(h^{-1}M_{\odot})$ &    \\
\hline 
\hline
Bower2006    & $1.08$ & $11.85$ & $11.72$ & $12.78$ & $0.276$ \\
DeLucia2007  & $1.01$ & $11.95$ & $11.74$ & $12.75$ & $0.114$  \\
\hline
\end{tabular}
\caption{
The best fitting parameters on applying the 5-parameter 
fit of Zheng et~al. (2005) to the HOD of stellar mass selected 
samples predicted by the Bower2006 and DeLucia2007 models, 
for the intermediate density cut.}
\label{table:5param} 
\end{center}
\end{table}

\begin{table*}
\begin{center}
\begin{tabular}{ | l | l | l | l | l | l | l | l | l | l |}
\hline
Model & $F_{\rm a}$ &$F_{\rm b}$         &$M_{\rm c}$         &$ (a\sigma_{\log M})$&$(b \sigma_{\log M})$&$F_{\rm s}$&$M_{\rm min}$           &$\delta_{\log M} $&$\alpha$ \\ 
      &             &                    &$(h^{-1}M_{\odot})$ &                     &                     &           &$(h^{-1}M_{\odot})$     &                         &         \\ 
\hline 
\hline
Cold Gas Mass \\
\hline
Bower2006  & $0$                   & $0.675$ & $11.36$ & $0.130$ & $0.395$ & $8.00\times 10^{-3}$ & $12.0$  & $0.586$ & $0.864$ \\
DeLucia2007 & $1.87 \times 10^{-2}$ & $0.97$  & $11.95$ & $0.34$  & $0.49$  & $3.7\times 10^{-3}$  & $11.3$  & $1.126$ & $0.854$ \\
\hline
SFR \\
\hline
Bower2006     & $6 \times 10^{-4}$    & $0.66$   & $11.53$ & $0.200$ & $0.380$ & $1.25\times 10^{-3}$ & $11.3$ & $0.987$   & $0.942$ \\
DeLucia2007     & $3.2 \times 10^{-2}$  & $0.92$   & $11.86$ & $0.308$ & $0.392$ & $8.75\times 10^{-4}$ & $10.5$ & $0.205$   & $0.895$ \\
  \hline
  \end{tabular}
  \caption{
The best fitting parameters when applying the 9-parameter 
fit (Eqns. A3 \& A4) to the HOD predicted by the Bower2006 and DeLucia2007 
models for intermediate density samples selected by their 
cold gas mass (top half) and star formation rate (bottom half). 
  }
\label{table:9param} 
\end{center}
\end{table*}

\begin{figure*}
\includegraphics[width=0.96\textwidth]{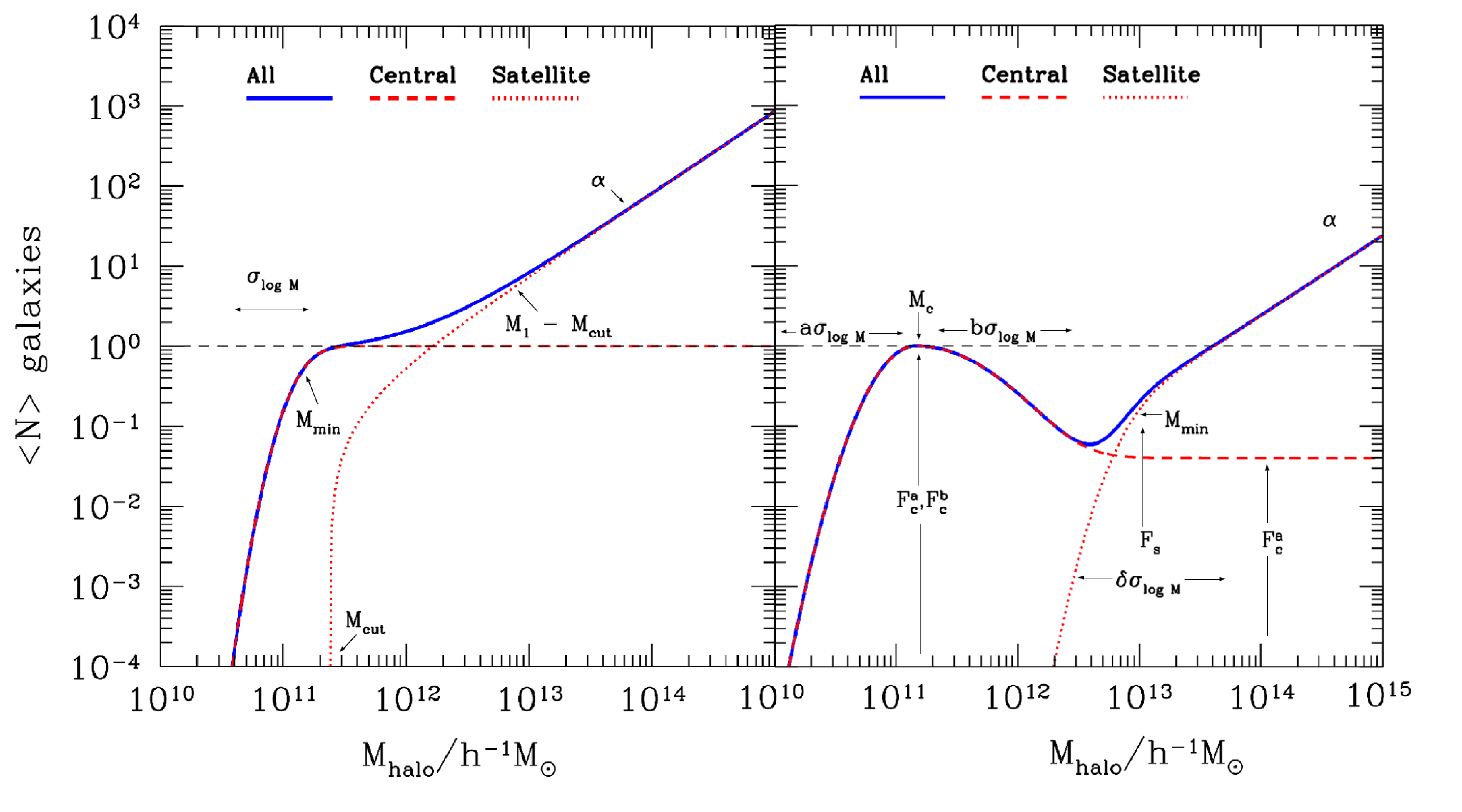}
\caption{
A schematic illustrating the parameteric forms used to fit the HODs 
predicted by the semi-analytical models. The overall HOD (blue) is 
separated into the contributions from central galaxies (red dashed) 
and satellites (red dotted). 
The left panel shows the 5-parameter fit of Zheng et~al. (2005).  
(Eqns. A1-A3) and the right panel shows the 9-parameter fit of 
Geach et~al. (2012)  (Eqns. A4 and A5). The labels give the 
parameter names as they appear in these equations and indicate 
which features of the HOD they control. 
}
\label{fig:HOD_DEF} 
\end{figure*}

\begin{figure*}
\includegraphics[width=0.43\textwidth, angle=270]{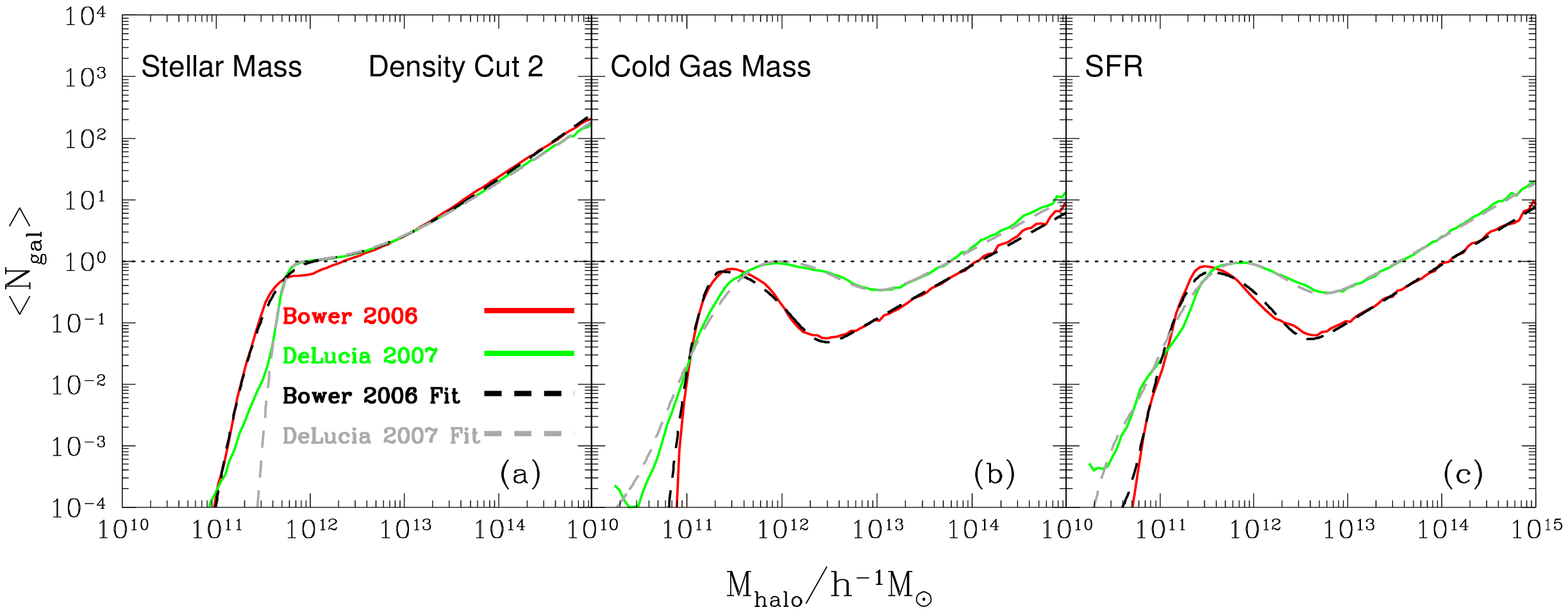}
\caption{
The solid lines show the HOD predicted by the Bower2006 and DeLucia2007 
semi-analytical models and the dashed lines show the best fitting parametric forms to these, with the parameters listed in Tables A1 and A2. 
Each panel shows the predicted HODs and the fits for a different ranking 
of galaxies, based on stellar mass (left panel), cold gas mass (middle) 
and star formation rate (right panel). In all cases, the intermediate 
density sample is shown. }
\label{fig:HOD_fit} 
\end{figure*}  
 
Semi-analytical galaxy formation models follow the 
processes which influence the baryonic component 
of the Universe, in order to predict the number of 
galaxies hosted by dark matter haloes along with 
the properties of the galaxies. The HOD is therefore 
a natural prediction of semi-analytical models. 
In this appendix, we compare the HODs output by the 
semi-analytical models with examples of the 
parametric forms typically used in the HOD analyses 
of galaxy clustering. Note that we fit the parametric 
form directly to the HOD output by the model and not 
to the HOD predictions for the correlation function 
and galaxy abundance, as is done when comparing 
HOD models to observations \citep{Cooray:2002,Zehavi:2005}. 

As we remarked in the paper, the form of the HOD predicted 
by the models is sensitive to the manner in which galaxies 
are selected. Following the standard practice with 
HOD fitting, we can separate the model output into contributions 
from central galaxies and satellites. The largest variations 
in the model predictions are seen in the HOD of central galaxies 
when different selections are applied. The central HOD for 
samples defined by stellar mass is almost a step function, 
but with a more gradual transition from a mean occupation 
of zero galaxies per halo to one galaxy per halo. 
For samples corresponding to high stellar mass galaxies, 
the predicted central galaxy HOD need not reach unity, 
even for the most massive haloes. In the case of samples 
constructed on the basis of star formation rate or cold 
gas mass, the central HOD is peaked rather than a step. 
Such a feature would be hard to anticipate but can be readily 
understood in terms of the physics in the semi-analytical models. 
The HOD of satellite galaxies is approximately a power law and 
varies mainly in normalization between different selections.   

The first parametrization of the HOD we consider is the five parameter 
form proposed by \cite{Zheng:2005}. The HOD is broken up into the  
contributions from the mean number of satellite galaxies 
$N_{\rm sat}(M)$ and centrals $N_{\rm cen}(M)$ (i.e. the fraction of 
halos of a given mass which contain a central galaxy passing 
the selection if $N_{\rm cen}(M)<1$) per halo, with the 
overall mean number of galaxies per halo 
given by $N(M) = N_{\rm cen}(M) + N_{\rm sat}(M)$.
The central galaxy HOD is a rounded step function, with a gradual 
transition from zero to one central galaxy per halo on average: 
\begin{equation}
 N_{\rm cen}(M) = \frac{1}{2}\left[ 1 + {\rm erf} \left( \frac{\log M - \log M_{min}}{\sigma_{\log M}}  \right) \right].
\end{equation}
Here $M$ is the mass of the host dark matter halo, 
$ {\rm erf}(x)$ the error function,
\begin{equation}
 {\rm erf}(x) = \frac{2}{\sqrt{\pi}} \int_{0}^{x} e^{-t^2} {\rm d}t,
\end{equation}
$M_{\rm min}$ is the mass at which the half of the 
halos have at least one galaxy and $\sigma_{\log M}$ 
is the width in halo mass of the transition from zero 
to one galaxy per halo. 
The HOD for satellite galaxies is given by 
\begin{equation}
 N_{\rm sat}(M) = \left( \frac{M-M_{\rm cut}}{M_1}\right)^\alpha, 
\end{equation}
where $M_{\rm cut}$ is the mass below which 
a halo can not host a satellite galaxy and $M_1$ 
is the mass in which a halo contains on average 1 satellite galaxy 
and $\alpha$ is the power-law slope, which usually has a 
value close to unity.

A schematic of the 5-parameter HOD with an explanation of 
which features the parameters control is given in the left-hand 
panel of Fig.~A1. The best fitting values of the parameters of 
this HOD model are listed in Table~A1 for the intermediate abundance 
stellar mass selected samples from the Bower2006 and DeLucia2007 
models. The predicted HODs and the fits are plotted in Fig.~A2. 

In the case of samples selected by cold gas mass or SFR, 
the form of the predicted HOD is very different from 
that for stellar mass selected samples. The HOD of central 
galaxies is peaked, as noted by \cite{Kim:2011}.
The five parameter HOD cannot describe this behaviour. 
Instead, the nine parameter fit advocated by \cite{Geach:2012} 
is more appropriate, with one modification. 
In this case, the central HOD is now given by  
\begin{eqnarray}
 N_{\rm cen}(M) = F_{\rm c}^B(1-F_{\rm c}^A) \exp \left[-\frac{\log(M/M_{\rm c})^2}{2 (x\sigma_{log M})^{2}} \right]+ \\ \nonumber 
F_{\rm c}^A \left[ 1 + {\rm erf} \left(  \frac{\log(M/M_{\rm c})}{x\sigma_{log M}} \right) \right]
\end{eqnarray}
where $F_c^{A,B}$ are normalization factors (with values varying 
between zero and one), $M_{\rm c}$ is the central value of the Gaussian. 
The modification we have made is to allow the width of the Gaussian 
to be different for masses on either side of $M_{\rm c}$.  
The dispersion is $x \sigma_{\log M} = a \sigma_{\log M}$ for halos with mass $M<M_{\rm c}$ and $b \sigma_{\log M}$ for those with $M>M_{\rm c}$.
This allows the best fitting central HOD to be an asymmetric peak. 
The satellite HOD is given by: 
 \begin{equation}
 N_{\rm sat}(M) = F_{\rm s} \left[  1 + {\rm erf} \left( \frac{\log(M/M_{\rm min})}{\delta_{\log M}}   \right) \right] \left(\frac{M}{M_{\rm min}} \right)^\alpha, 
 \end{equation} 
where $F_{\rm s}$ is the mean number of satellites per halo halo at 
the ``transition'' mass $M_{\rm min}$, which corresponds roughly to 
the mass above which the mean number of galaxies per halo is dominated 
by satellites, $\delta_{\log M}$ represents the width of the 
transition from zero satellites per halo to the power law,  
and $\alpha$ is the slope of the power-law which gives the mean 
number of satellites for $M > M_{\rm min}$ 

A schematic of the 9-parameter HOD, along with an explanation of 
which features the parameters control is given in the right-hand 
panel of Fig.~A1. The best fitting values of the parameters of 
this HOD model are listed in Table~A2 for the intermediate abundance 
cold gas mass and star formation rate selected samples from 
the Bower2006 and DeLucia2007 models. The predicted HODs 
and the fits are plotted in Fig.~A2.

\end{document}